\def\asc{$^{\prime\prime}$}
\documentclass[preprint]{aastex}

\shorttitle{Palomar 8 pc Survey}
\shortauthors{Oppenheimer et al.}
\begin{document}

\title{Coronagraphic Survey for Companions of Stars within 8 pc}

\author{B. R. Oppenheimer\altaffilmark{1}}
\affil{Palomar Observatory, 105-24 California Institute of Technology, 
Pasadena, CA 91125 USA}
\altaffiltext{1}{Present Address: Astronomy Department, University
of California, Berkeley, 601 Campbell Hall MS3411, CA 94720, USA}
\email{bro@astron.berkeley.edu}

\author{D. A. Golimowski}
\affil{Department of Physics and Astronomy, The Johns Hopkins University,
Baltimore, MD 21218 USA}

\author{S. R. Kulkarni, K. Matthews, T. Nakajima\altaffilmark{2}, M. Creech-Eakman}
\affil{Palomar Observatory, 105-24 California Institute of Technology, 
Pasadena, CA 91125 USA}
\altaffiltext{2}{Present Address: National Astronomical Observatory, Tenmondai, 2-21-1, Oshawa, Mitaka, Tokyo, Japan}

\and

\author{S. T. Durrance\altaffilmark{3}}
\affil{Department of Physics and Astronomy, The Johns Hopkins University,
Baltimore, MD 21218 USA}
\altaffiltext{3}{Present Address: Florida Space Institute, MS:FSI, Kennedy Space Center, FL 32899}

\begin{abstract}
We present the technique and results of a survey of stars within 8 pc
of the Sun with declinations $\bf\delta > -35\arcdeg$ (J2000.00).  The
survey, designed to find without color bias faint companions, consists
of optical coronagraphic images of the 1\arcmin\ field of view
centered on each star and infrared direct images with a 32\asc\ field
of view.  The images were obtained through the optical Gunn $r$ and
$z$ filters and the infrared J and K filters.  The survey achieves
sensitivities up to four absolute magnitudes fainter than the
prototype brown dwarf, Gliese 229B.  However, this sensitivity varies
with the seeing conditions, the intrinsic brightness of the star
observed and the angular distance from the star.  As a result we
tabulate sensitivity limits for each star in the survey.  We used the
criterion of common proper motion to distinguish companions and to
determine their luminosities.  In addition to the brown dwarf Gliese
229B, we have identified 6 new stellar companions of the sample stars.
Since the survey began, accurate trigonometric parallax measurements
for most of the stars have become available.  As a result some of the
stars we originally included should no longer be included in the 8 pc
sample.  In addition, the 8 pc sample is incomplete at the faint end
of the main sequence complicating our calculation of the binary
fraction of brown dwarfs.  We assess the sensitivity of the survey to
stellar companions and to brown dwarf companions of various masses and
ages.
\end{abstract}


\section{Introduction}

In 1992 a brown dwarf companion search began at Palomar with the
initiation of a collaboration between T. Nakajima and S. Kulkarni at
Caltech and D. Golimowski and S. Durrance at Johns Hopkins.  The
Hopkins group brought the Adaptive Optics Coronagraph (AOC;
\citet{G92}) to Palomar to be fitted on the 60'' Telescope.  The first
results of this collaboration are given in \citet{N94}, which entailed
a search for companions of high galactic latitude stars.  Companions
were distinguished from background stars through statistical arguments
based on the distribution of point sources as a function of angular
separation from the stars.  In 1994, the work expanded to a new sample
of nearby stars which were believed to be young.  A short description
of this sample is contained in \citet{N95}.  This sample was biased
toward young stars in an attempt to discover brown dwarf companions,
with the assumption that younger brown dwarfs would be easier to
detect because they should be brighter \citep{O00}.  The first success of
this collaboration was the discovery of a faint companion of the star
Gliese 105A \citep{gol95}.  The first success of the young
star survey was the discovery of the cool brown dwarf Gliese 229B
\citep{N95,O95}.

Following the discovery of Gliese 229B, we decided that it was
of paramount importance to conduct a volume-limited survey for
companions.  If we continued to pursue the biased sample and found no
more brown dwarfs, we would have little to say about the prevalence of
companion brown dwarfs without extensive modeling.

For this reason, in late 1994 we began a survey of all the Northern
($\delta > -35\arcdeg$) stars within 8 pc of the Sun to search for
brown dwarf companions.  Because all of the known stars within 8 pc
have measurable proper motions, our survey was designed to find common
proper motion companions.  We thus observed each star at multiple
epochs, if point sources other than the star appeared in the field of
view.  This permitted us to discern companions simply by measuring the
relative offset between the star and the putative companions at each
epoch.  The common proper motion criterion, almost fifty years old
now, is ideal in searches for brown dwarfs because it is intrinsically
unbiased by color or other theoretical notions of what a brown dwarf
should look like.  (Other systematic searches for companions that used
the common proper motion criterion are by
\citet{vb61,luyt,skrut89,sim96,koern99} and \citet{schr00}. See
\citet{O00} for a description of the history of brown dwarf searches.)
The common proper motion criterion is the most physically rigorous
short-term method for finding companions.  (Longer term methods
include orbital motion measurements and common parallax measurements,
which eliminate the minute possibility that two objects within an
arcminute of each other might exhibit common proper motion and yet be
physically unassociated.)

This survey is also distinguished from others because it represents
the first use of adaptive optics techniques in the study of nearby
stars.  With our tip-tilt observations dating back to 1992, we greatly
predate any other such searches.  At the time of this writing, the use
of higher order adaptive optics systems is becoming widespread in
these sorts of studies (e.g., \citet{delfoss98b}).  The combination of
adaptive optics and coronagraphy, and the use of both infrared and
optical band passes made our search effective.  We demonstrate in \S 9
and Fig. 19 that the survey covers previously unobserved parts of the
companion mass-separation parameter space.

To achieve our goal of a volume-limited survey, we assembled a sample
of stars which appear in the Third Catalog of Nearby Stars
\citep{GJ91} and have parallaxes greater than 0\farcs125.  The
degree of completeness of this sample has been the subject of debate.
(See, for example, \citet{reid97}.)  We address this issue in depth in
\S \ref{chap4:8 pc} where we also present an updated catalog of the stars
within 8 pc.

The observations are explained in detail in \S \ref{chap4:obs}, and a
complete description of the sensitivity limits is presented in later
sections. 

\section{The 8 pc Sample}\label{chap4:8 pc}

\subsection{Culling the Catalog}

When we began our survey the best list of stars within 8 pc of the Sun
was a subset of the Third Catalog of Nearby Stars (\citet{GJ91};
CNS3).  The subsequent releases of the Hipparcos Main Catalog
\citep{esa97} and the Yale Catalog of Trigonometric Parallaxes
\citep{yale95} provide important sets of data that modify the census
of stars within 8 pc.  Ultimately the two new catalogs moved some
stars out of, and others into, the 8 pc sample.  These catalogs also
added a few stars to the sample, which were not in the CNS3.  The
Hipparcos Catalog did not add any unknown stars to the 8 pc sample
because trigonometric parallaxes were only obtained for previously
cataloged stars.  However, Hipparcos did measure 7 stars in 5 systems
whose parallaxes had never before been measured and which place them
within 8 pc.  The Yale Catalog adds no new stars to the catalog but
does provide trigonometric parallaxes for 37 stars within 8 pc which
were not measured by Hipparcos.  This reduces the number of stars in
our sample whose parallaxes are simply inferred from photometry.
These so-called photometric parallaxes involve measurements of the
colors of a given star.  The colors determine the spectral class of
the star and, thus, its absolute luminosity.  From the absolute
luminosity, the distance modulus is calculated.  These photometric
parallaxes are sometimes inaccurate because of unknown multiplicity
and intrinsic scatter in the main sequence which can lead to errors in
the distance as large as 30\% \citep{weis84}.  In the new catalog we
assemble here, only 6 stars are included based on photometric
parallaxes.  These are the only stars within 8 pc in the CNS3 which
were not measured astrometrically by the Hipparcos or the Yale
surveys.

Pursuant to this discussion we have created a new catalog of stars
within 8 pc of the Sun.  We believe this comprises the most complete
census of the Northern 8 pc volume to date.  In our catalog we combine
all the stars within the CNS3, Hipparcos and Yale catalogs which have
parallaxes greater than 0\farcs125.  The three catalogs are fully
cross-correlated and for each entry in our database we have up to
three different parallaxes, although only the most accurate is listed
in the catalog presented here.  Precedence for inclusion in the final
catalog is given to the Hipparcos measurement which is generally more
accurate than the Yale measurement (except in the case of star systems
Gliese 185AB and Gliese 644ABCD, where the Yale parallax is more
accurate).  For six of the stars neither Yale nor Hipparcos
measurements exist.  These stars have photometric parallaxes listed in
the CNS3 and we list them to be as inclusive as possible.

To be certain that we have included all the known subordinate stellar
and substellar objects associated with these stars, we conducted a
search of the literature.  Two papers in particular
\citep{delfoss98b,reid97} provided new companions, some of which have
been resolved and some of which were detected through radial velocity
studies.  Our catalog is not biased in any way as to whether a given
companion has been visually resolved.  A complete census must be free
of these considerations.  Therefore, we include all of the companions
mentioned in \citet{reid97} and most of those in \citet{delfoss98b}.
We believe that our final catalog is complete as of January 2000
because we have used all of the available resources and studies of the
nearby stars.

Our sample includes, therefore, 163 stars, two brown dwarfs
\citep{N95,bur00} and one indirectly detected planet \citep{delfoss98a}.
These entities are arranged in 111 star systems, 29 of which are
double, nine of which are triple, two of which is quadruple and one of
which is quintuple.  

Table \ref{chap4:tab1} lists all the star systems in the 8 pc sample
described above.  In this table we give a single entry for every
object known within 8 pc.  For multiple systems, entries are grouped
together and indicate the separation of the subordinate components,
along with other vital data.  The table is arranged in order of
decreasing parallax in milliarcseconds (mas).  The ``Source Code''
field in each entry indicates where the parallax measurement comes
from.  Companions of stars are indicated by capital Roman letters
after the parallax and generally are given in the order in which the
components are discovered.  An implicit A is given to every principal
star in each star system.  However, the A is only used if at least
component B has been discovered.  For convenience we give HD,
Durchmusterung, CNS3 or other names for the stars if they are
available.  This permits easy identification of the stars in
astronomical databases.

\subsection{Completeness of the 8 pc Catalog}

It is important before describing the observations we undertook to
estimate how complete our catalog of star systems within 8 pc is.  

The simplest way to assess this involves extrapolating the number of
stars within 5 pc to the volume of the 8 pc sample.  This sort of
analysis was conducted by \citet{hen97}.  In their estimation,
the CNS3 is complete for stars with $M_V < 11$ all the way out to 10
pc, although the majority of the incompleteness is in the
far-less-studied Southern sky ($\delta < -35^\circ$).  Their
computation involves taking the densities of stars of various absolute
magnitudes in the 5 pc volume in the CNS3 (widely used as a benchmark
for complete stellar samples) and multiplying by the ratio of the
volumes due to increasing the radius of the sample.  The 5 pc sample
in the CNS3 contains 53 stars with $\delta > -35^\circ$.  This implies
that there should be another $3.096 \times 53 \pm 7.3 = 164 \pm 23$
stars between 5 and 8 pc.  However, there are only 110 such stars
known.  Since we surveyed 163 stars but expect those to be drawn from
a population of 217 $\pm$ 30, we estimate that our
catalog is complete to approximately 75\%\ (assuming the stellar
densities within 5 pc are correct).  Any incompleteness is most likely
among the very faintest stars (white dwarfs and late M dwarfs),
because they are less likely to have been measured and studied in
depth in large scale stellar surveys.  In addition, many of the nearby
stars were found through large-scale proper motion surveys.  Some
stars may therefore be missing from the nearby star sample because
they have very small proper motions.

\citet{reid97} argue that the CNS3 is complete to $M_V <
14$ for $\delta > -30^\circ$ within 10 pc.  $M_V = 14$ corresponds to
the spectral type M4.5.  These considerations permit us to conclude
that our sample is complete at least to the spectral type M5, and
possibly even fainter.  We conservatively claim that the sample is
75\%\ complete and believe that the missing stars are all later than
M5 in spectral type.

\section{Observations}\label{chap4:obs}

In pursuit of our goal to image and study brown dwarf companions of
stars in our sample, we conducted observations of 107 of the 111 star
systems (96\%\ of the sample) in optical and near infrared
wavelengths.  The observations employed two different imaging
instruments, the AOC attached to the Palomar 60'' Telescope and the
Cassegrain Infrared Camera fitted to the Palomar 200'' Hale Telescope.

\subsection{Common Proper Motion}

We imaged each of the stars at least twice in order to discern faint
objects within 30\asc\ of each star which display the same proper
motion as the stars themselves.  Our decision to use the common proper
motion criterion was motivated by the plethora of models of brown
dwarfs at the time the survey began.  These models painted often
conflicting depictions of the colors or spectra that brown dwarfs
ought to exhibit.  (See \citet{BL93} for a comprehensive review of the
state of these models at about the time that our survey began.)  We
decided that instead of relying upon one of the models and designing a
survey that looked for colors that such a model predicted, we would
use a more basic physical argument for finding cool companions. 

\begin{figure}
\plotone{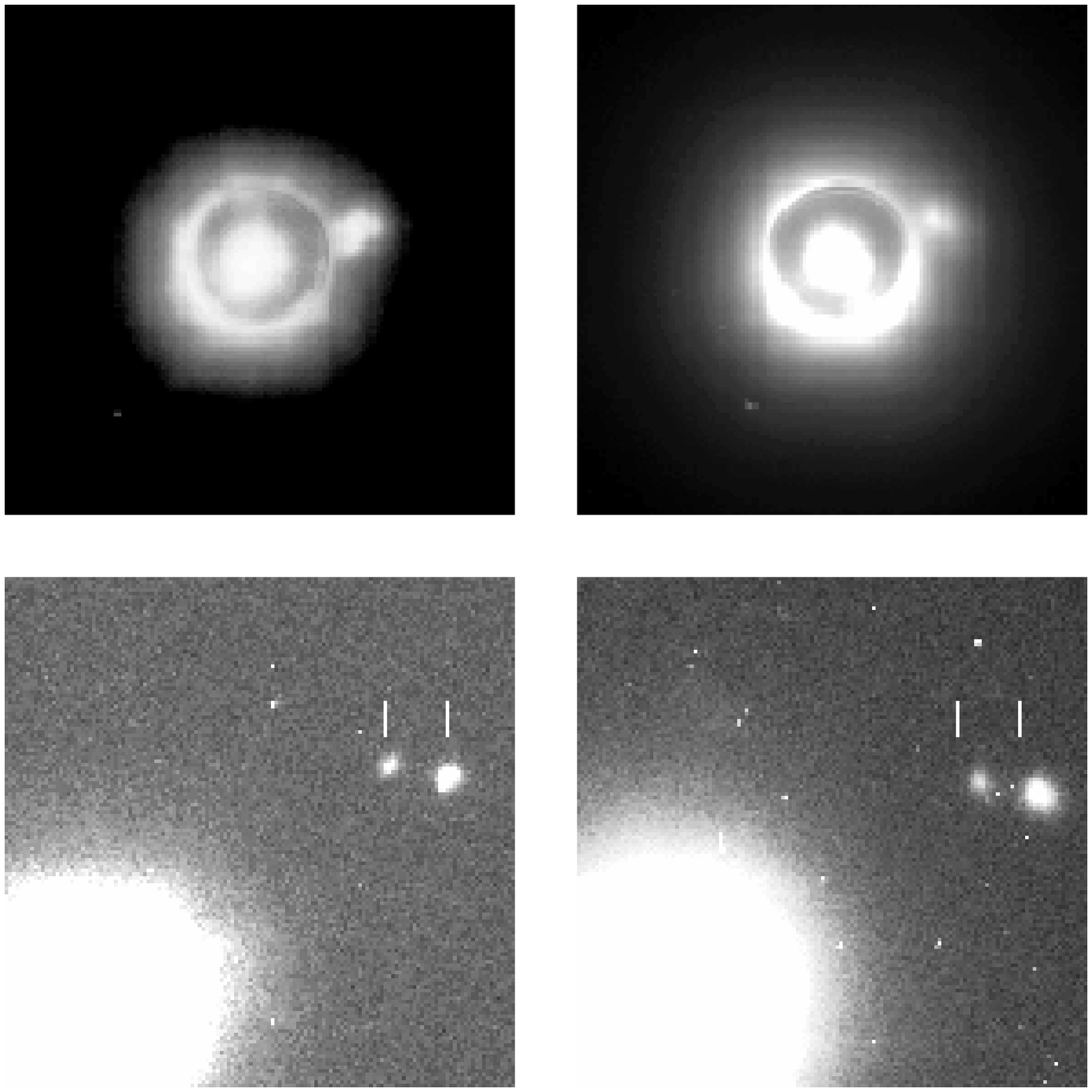}
\caption{Images of Gliese 105AC (138.72AC).
The top two panels are magnified portions of the lower panels.  These
images were taken in the $z$ band in October 1994 (left) and October
1995 (right).  North is up and East is to the left.  The star has
moved over 2\farcs2 between these two epochs and yet the fainter
object maintains the same offset from the star in the top panels.
This makes it a common proper motion companion.  The seeing was better
than 0\farcs6 in the images, and the astigmatism (which has since been
fixed) of the 60'' telescope is apparent.  The pixel size is
0\farcs117.  The top images measure 16\asc\ on a side.  The lower
images are 40\asc\ on a side.  The two white tick marks in the lower
left panel serve to mark the positions of the two field stars in 1994.
In the lower right panel they are placed in the same location to
clearly show that the field stars have moved relative to the central
star (which is placed in the same location in the two panels).
\label{fig:c4f1}}
\end{figure}

The technique we employ here is explained through example in Fig.\
\ref{fig:c4f1}.  This figure shows four $z$ band coronagraphic images
of the star Gliese 105A (138.72AC).  The first two images are
magnified portions of the region immediately around the star.  These
two images were taken a year apart and show a faint second object with
the same proper motion.  The star moves 2\farcs2 yr$^{-1}$ permitting
extremely easy discrimination between common proper motion companions
and background stars.  This is demonstrated in the bottom two images,
which are larger portions of the same images.  In these two, one
can clearly see that the two stars to the West do not share the proper
motion of the star.  We reported this common proper motion companion
in the paper by \citet{gol95}.

The error in the measurement of the centroid of a stellar image on a
CCD is approximately the angular size of the image divided by the
signal-to-noise ratio.  (There is a correction factor if the pixel
size is much smaller than the image size, but this amounts to a 0.6\%\
correction in this survey.  This is because we require every star to
be imaged in 1\asc\ seeing or better.)  In our survey we combine $n$
measurements of each star to improve upon this standard astrometric
limit.  The effective signal-to-noise ratio of the combined
measurement is improved by the factor $n^{1/2}$.  At each epoch we
have at least two images.  With at least two epochs per star, the
value of $n$ for most stars is greater than four.  The average value
of $n$ for all the stars in the survey with other point sources in the
field of view is eight.  This means that the average centroid error
for sources detected at the 5 $\sigma$ level is 0\farcs07.  All of the
stars in our sample exhibit proper motions greater than 0\farcs07
yr$^{-1}$.  Thus we are capable of discerning background objects from
common proper motion companions with these observations.  In most
cases, the proper motions are actually substantially larger than
0\farcs07 yr$^{-1}$ and the requirement on the astrometric errors is
far less stringent than 0\farcs07.

\subsection{Optical Observations}

Optical coronagraphic images of the survey stars were obtained during
25 separate observing runs on the Palomar 60'' Telescope between
September 1992 and April 1999.  These images were obtained with a
Tektronix 1024 $\times$ 1024 pixel CCD camera binned in a 2 $\times$ 2
pattern (0\farcs117 pixel$^{-1}$) attached to the back-end of the AOC.
This device consists of a standard Lyot coronagraph \citep{lyot39}
fitted behind a tip-tilt mirror which uses the occulted star as a
guide star.  The tip-tilt correction provides substantial gains in
image resolution on the Palomar 60'' Telescope because, there, the
majority of the atmospheric disruption of the stellar wavefront
resides in tip-tilt energy or, equivalently, image motion.  We
routinely obtained images with resolutions of 0\farcs7 and on six of
the observing runs we obtained images at 0\farcs45 resolution.  The
images were taken through the Gunn $r$ and $z$ filters, with
additional images taken through the $i$ band if a companion was found.
In almost all cases the CCD was exposed for 1000 s.  However, for the
stars whose magnitude is $V < 6$, we were forced to take shorter
exposures and to sum these to produce a final image with a 1000 s
exposure time.  The AOC on the 60'' telescope was incapable of guiding
on stars fainter than $V \sim 13.5$.  For this reason we were unable
to observe 13 of the sample stars with the AOC.  We had to rely upon
the infrared observations of these stars to discern companions.  These
stars are indicated by the words ``too faint'' in Table
\ref{chap4:tab2}.

The AOC's focal plane aluminum occulting stop is uniformly translucent
in the $r$ and $z$ bands.  This permits an accurate measurement of the
position of the star in the CCD image.  (Without the transparent stop,
pinpointing the star would have been impossible because the pupil
plane stop eliminates the diffraction spikes of the star.)  We used
coronagraphic stops 4\farcs2 in diameter for most observations, except
in the case of the brightest stars where an 8\farcs4 diameter stop was
used.  We claim no sensitivity to faint companions under the stops.
However, equal brightness binaries were sometimes resolved under the
masks (e.g., Fig.\ \ref{fig:089}).

If a set of $r$ and $z$ band images failed to reveal any sources in
the field of view other than the star, we would not reobserve the
star.  Table \ref{chap4:tab2} lists the dates of all the observations
of each star in the sample.  In all cases we attempted to acquire
images with seeing better than 1\farcs0.  Seeing worse than this
strongly degraded our sensitivity (\S \ref{chap4:sens}).  In all cases
we were able to obtain such images at least once for each of the stars
observed, and at least twice for those stars with possible companions
(i.e., with any other point source in the field of view).

Data reduction involved the subtraction of a bias frame from each of
the science images, division by a flat field image obtained using the
60'' dome and a flat field lamp, and the removal of cosmic rays (easy
to identify in these images because of the very small plate scale).
The images were then inspected by eye.  In most cases this was
sufficient to ascertain whether common proper motion companions were
present.  However, for the stars whose proper motions are small,
additional work was required to distinguish field stars from common
proper motion objects.  Because the central occulting mask of the
coronagraph was somewhat transparent, we were able to centroid the
light of the survey star to ascertain its position in the images.
Simple centroiding of the other point sources yielded pixel locations
as well.  These were converted into angular separations in arcseconds
by using the observing run's plate scale and detector orientation.
These were determined with the astrometric calibrator fields as
explained in \S \ref{sec:astrom}.

\subsection{Infrared Observations}

Direct infrared images of each of the sample stars were obtained with
the D78 Cassegrain Infrared Camera on the Palomar 200'' Telescope
during 15 observing runs between May 1995 and March 1999.  We allowed
the stars to saturate the central part of the 256 $\times$ 256 InSb
array (0\farcs125 pixel$^{-1}$).  The field of view in these images
was approximately 32\farcs\ ~Our imaging technique
entailed the use of the J and K filters with two types of exposure
through each filter.  The first type involved a total of five coadds
of 1 s exposure time.  These images were meant to reveal close
binaries with a dynamic range on the order of five to eight magnitudes
(depending on the seeing).  The second type of exposure used five
coadds of 10 s.  These exposures were designed to detect fainter
companions outside of a 3\asc\ radius from the star.  In each case a
sky frame was taken immediately after the data image was acquired.
The sky frame was taken in the same manner as the data image, but with
the telescope pointed 50\asc\ to the N or E.  In some cases we changed
this distance because of the presence of a rather bright source in the
sky frame.

As with the optical observations, if no objects other than the star
appeared, we would not observe the star a second time.  In addition
the seeing requirement of $\leq$ 1\farcs0 for the infrared
observations was identical to that for the optical images (see above).

The data reduction for the infrared images involved the subtraction of
the sky frame from the ``on-source'' frame.  Subsequent division by a
flat field (acquired from the twilight sky during each observing run)
permitted the more detailed examination of the images.  As with the
optical data, visual inspection of the images was usually enough to
discern common proper motion companions.  In cases where more accurate
measurements were necessary, we needed to pinpoint the location of the
survey star.  This could not be done through centroiding because all
of the stars's images were saturated.  Fortunately, we could use the
diffraction spikes in the infrared images (absent in the optical
images because of the pupil plane apodizing mask in the coronagraph).
By fitting the unsaturated parts of the two diffraction spikes with
perpendicular lines, we were able to localize the star with an
accuracy of better than a third of a pixel.  (We confirmed this
through short unsaturated exposures of some of the fainter stars in
the sample while the telescope guided on a field star.  The low
frequency tip-tilt system on the f/70 secondary mirror of the 200''
telescope guides with an accuracy of better than 0\farcs03 over 20
min.)  This stellar position on the detector was then used, along with
centroids of the light from putative companions, to compute offsets
between the objects.  These were converted into angular offsets in
arcseconds by using the plate scale as described in \S
\ref{sec:astrom}.

\subsection{Astrometric Calibration}\label{sec:astrom}

In order to obtain the accurate astrometric measurements of the
offsets between the central star and its putative companions, we made
observations of calibration fields during each observing run.  These
fields contained six to ten stars whose relative positions are known
within a few milliarcseconds.  We used these fields to determine that
the astrometric distortion on the face of the CCD chip used in the AOC
observations is smaller than 0\farcs01 over the whole chip.  In the
infrared observations the distortion is almost a full pixel near the
edges of the array.  This distortion is constant with time.  Because
we centered the stars in the same position on the infrared array at
each observing run, the comparison of astrometric measurements is
valid, despite this image distortion.  We did not use relative
astrometry measured on the infrared images in conjunction with other
measurements made on the AOC images.  For these reasons we applied no
astrometric distortion correction in our measurements of relative
offsets between stars.

For all of the infrared observing runs, the data were taken with the
Cassegrain ring angle precisely set to place North up and East left on
the array.  This prevented complications in astrometric measurements
that would arise from using arbitrary position angles of the array
with respect to the cardinal directions.

\begin{figure}
\plotone{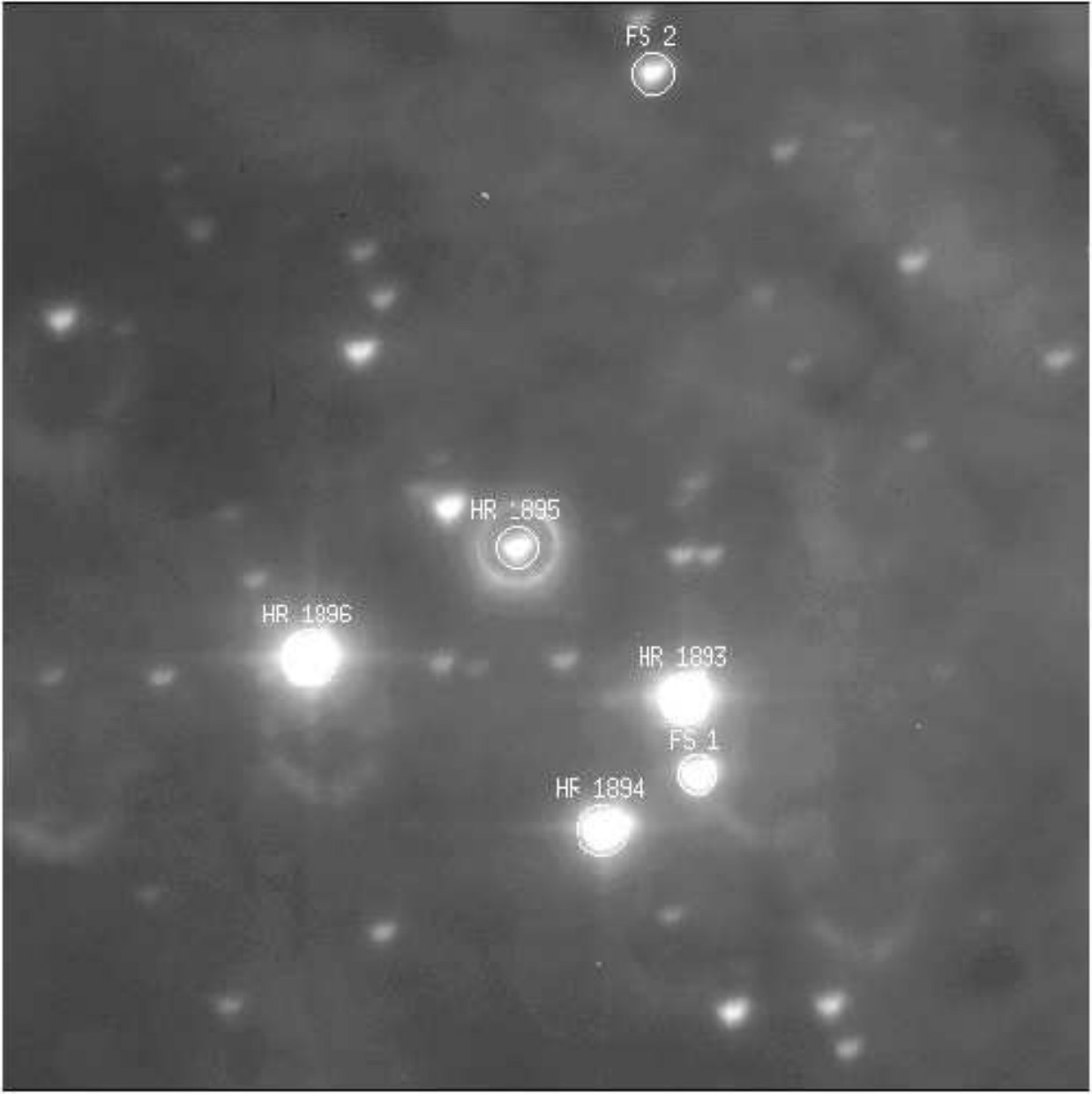}
\caption{The Trapezium
astrometric calibration field.  The stars marked with circles and
numbers are the stars listed in Table \ref{chap4:tab4}
and are the ones used to conduct the calibration.
The image measures one arcminute on a side.  North is down and East
left.\label{fig:astrofieldtrap}}
\end{figure}

\begin{figure}
\plotone{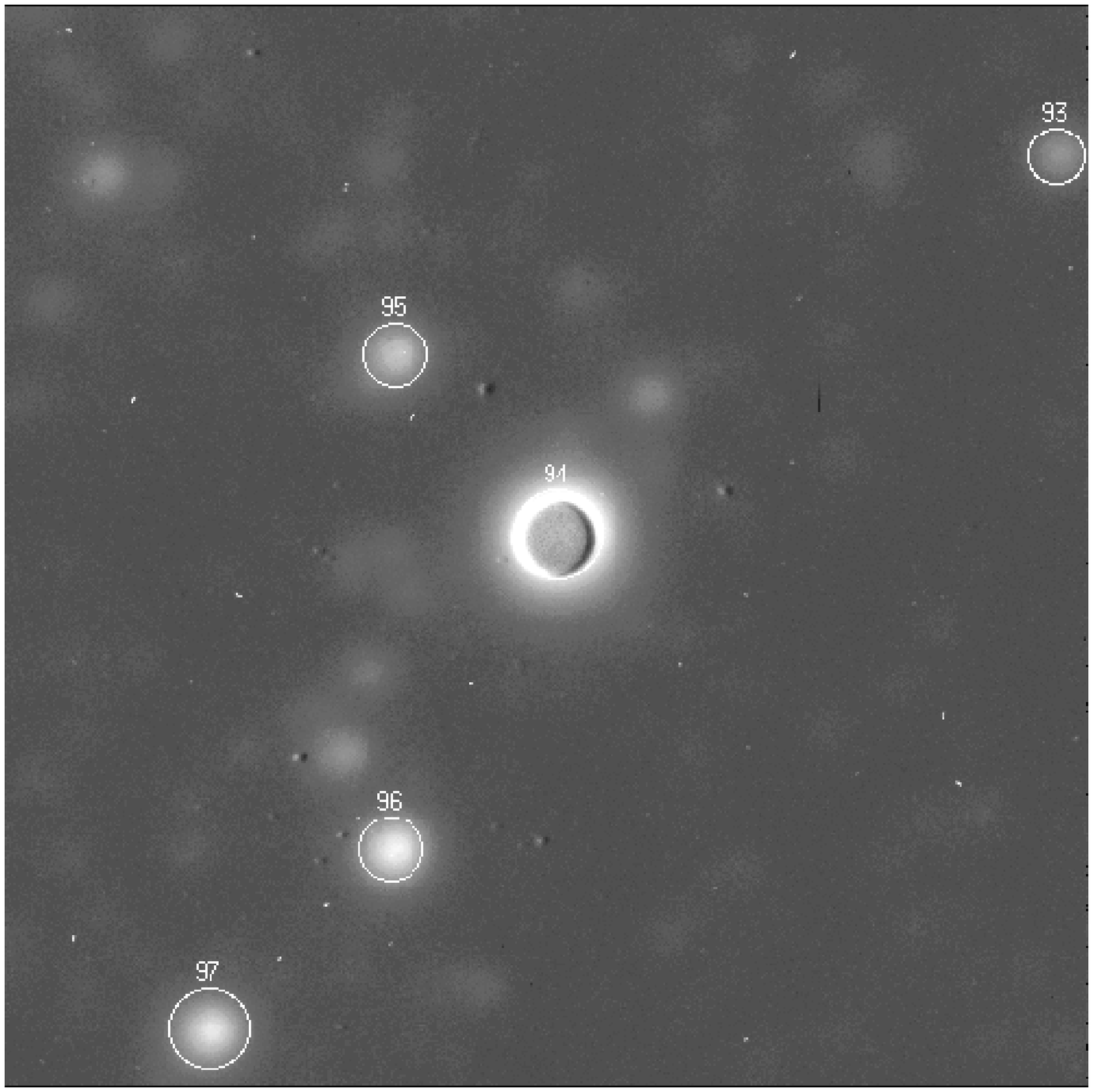}
\caption{The M5 astrometric
calibration field.  The stars marked with circles and numbers are the
stars listed in Table \ref{chap4:tab5} and are the ones used to
conduct the calibration.  The images measure one arcminute on a side.
North is left and East down.\label{fig:astrofieldm5}}
\end{figure}

\begin{figure}
\plotone{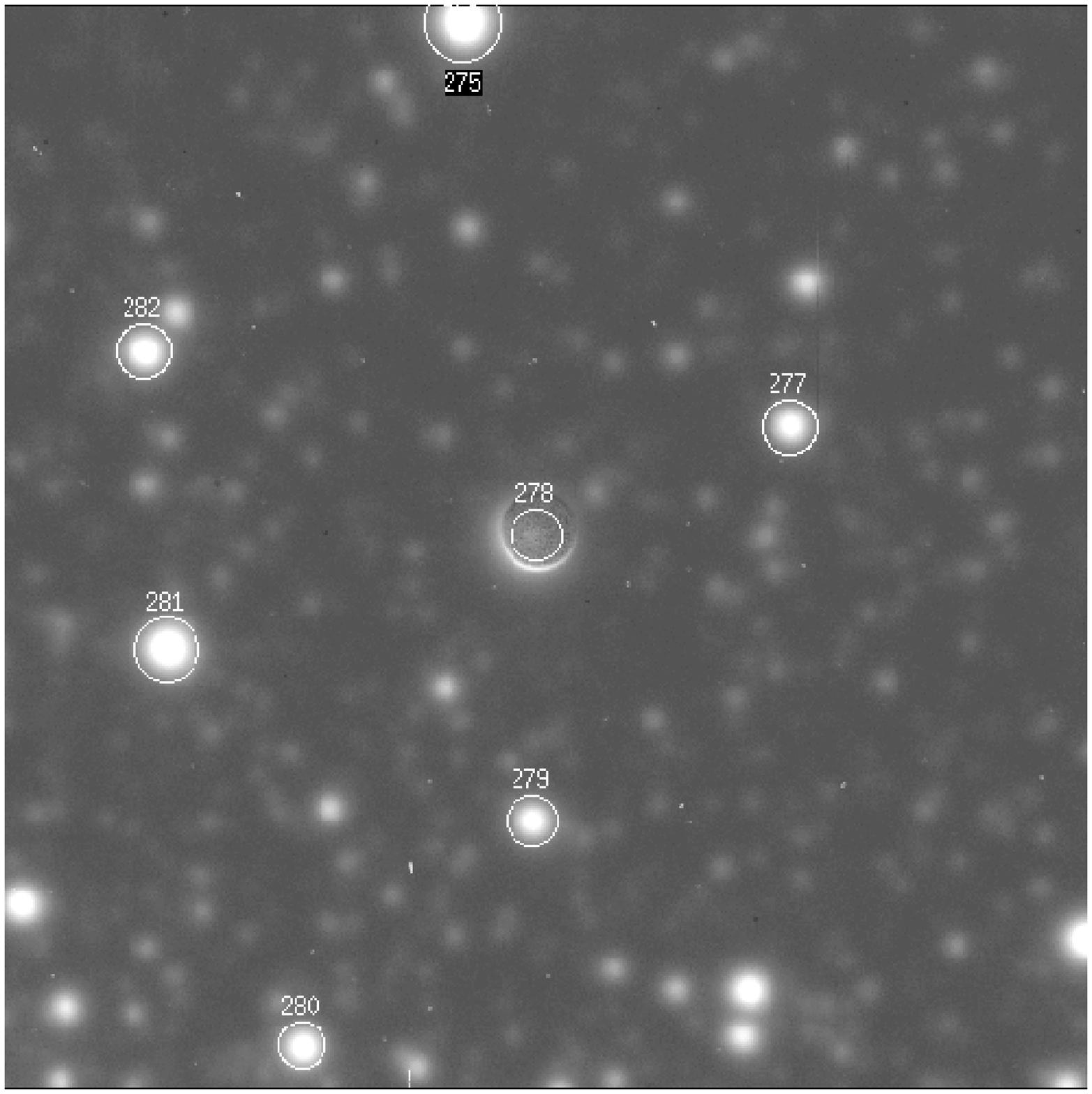}
\caption{The M15 astrometric
calibration field.  The stars marked with circles and numbers are the
stars listed in Table \ref{chap4:tab6} and are the ones used to
conduct the calibration.  The image measures one arcminute on a side.
North is left and East down.\label{fig:astrofieldm15}}
\end{figure}

For every AOC observing run we would observe whichever of the three
astrometric calibration fields (listed in Tables \ref{chap4:tab4},
\ref{chap4:tab5} and \ref{chap4:tab6}) was visible.  In each case we would
place the star shown in the center of each of Figs.\
\ref{fig:astrofieldtrap}, \ref{fig:astrofieldm5} and 
\ref{fig:astrofieldm15} in the center of the field of view.  This star
was used for guiding and rapid tip-tilt image motion compensation.
From the images---500 s $r$ band exposures---we were able to
determine precisely the plate scale and the rotation of the CCD on
the plane of the sky.  We found that the plate scale was extremely
stable in both instruments (despite the fact that the CCD camera used
on the AOC was taken apart and reassembled twice between the starting
and ending dates of the survey).  The accurate positions of the stars
in these fields are from \citet{cud79} for M5, \citet{cud76} for
M15 and \citet{mcst94} for the Trapezium.

Figs.\ \ref{fig:astrofieldtrap}, \ref{fig:astrofieldm5} and
\ref{fig:astrofieldm15} show each of the three calibration fields and
Tables \ref{chap4:tab4}, \ref{chap4:tab5} and
\ref{chap4:tab6} give the positions of the stars used for the
astrometric calibration.

\section{Detection Limits for Each Star}\label{chap4:sens}

Typically in imaging surveys assessing the sensitivity of the images
and the survey as a whole simply involves a determination of the
limiting magnitudes of the images.  However, in this case the problem
is somewhat more complicated.  

\begin{figure}
\plotone{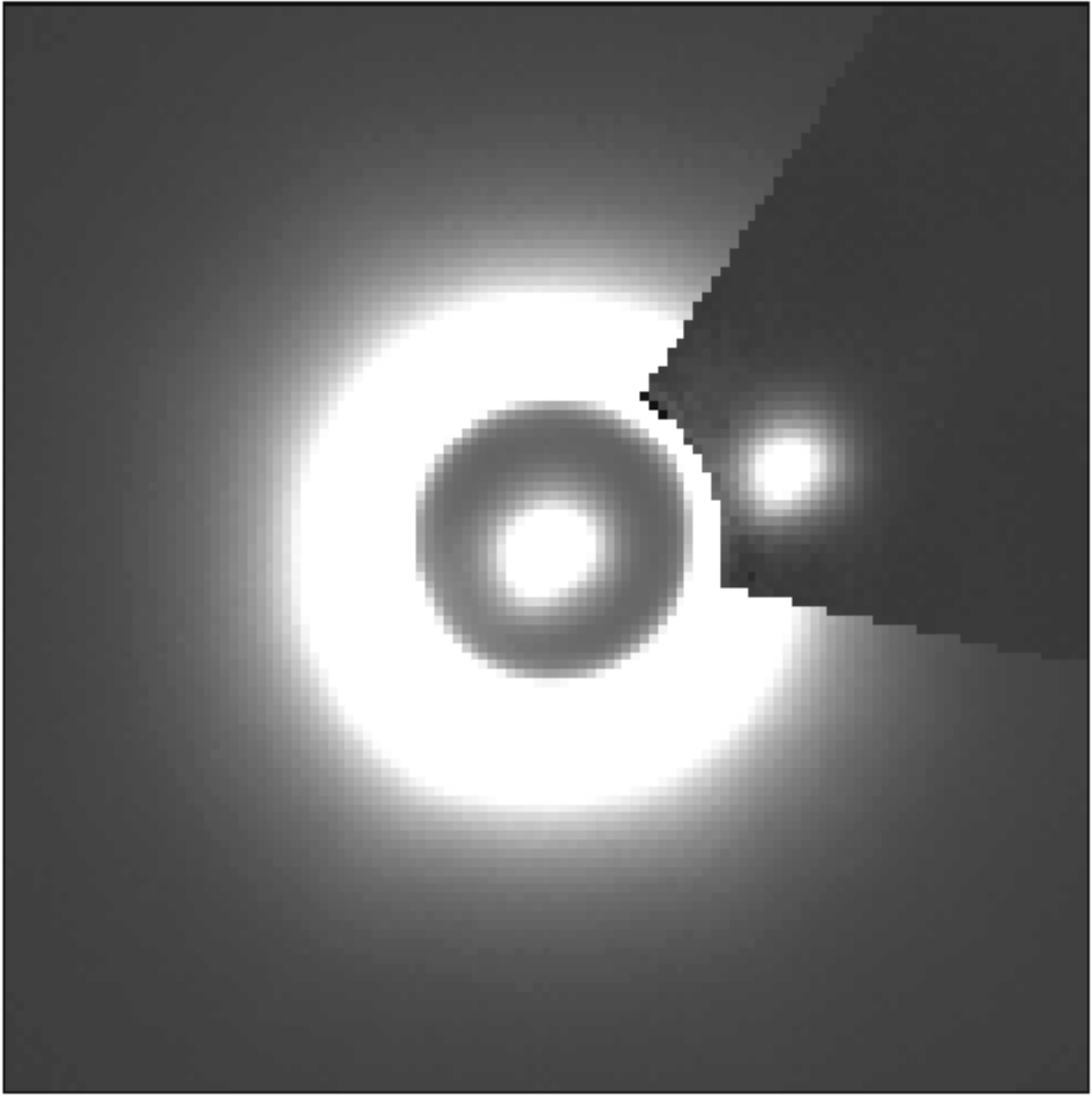}
\caption{Image of 
Gliese 105AC.  This image, taken through the $i$ band in October 1993,
shows only the inner 14\farcs5 $\times$ 14\farcs5 piece of the larger
AOC image.  North is up and East to the left.  The occulting mask
(somewhat transparent) is 4\farcs3 in diameter and reveals the core of
the star's seeing disk.  Outside the occulted region, part of the
seeing disk has been modeled and subtracted to make the companion
stand out better.  This is unnecessary in order to see the companion,
however, as shown in Fig.\ \ref{fig:c4f1}.  This image demonstrates
the huge dynamic range possible with the AOC.  Only 3\farcs3 from a
star with $i = 7.03^m$ we detected, with ease, a very low-mass star
with $i = 12.6^m$ with a signal-to-noise ratio of several thousand.
We could have detected a companion at this separation in this image as
faint as 18.5$^m$, indicating a dynamic range of 11.5$^m$ at 3\farcs3.
(From Golimowksi et al.\ 1995.)\label{fig:c4f2}}
\end{figure}

The presence of the bright star in our images means that over much of
the field of view the sensitivity is limited by the light of the star
and not the sky background (or read noise as was the case for speckle
interferometric surveys of these stars; \cite{hm90}).  However, this
survey extends into uncharted parameter space because the
coronagraphic technique suppresses a substantial portion of the
starlight.  The image shown in Fig.\ \ref{fig:c4f2} illustrates this
with the Gliese 105 AC system. 

\begin{figure}
\plotone{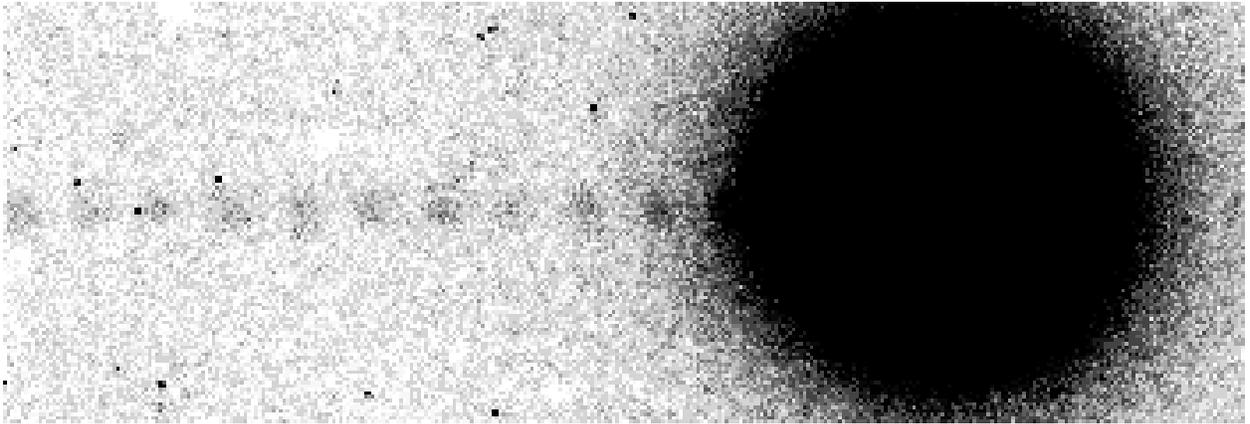}
\caption{Example of the
sensitivity curve determination technique.  The images shown are part
of a $z$ band coronagraphic image of Giclas 192$-$013 (132.10).  The
artificial point sources are visible to the left of the star and are
placed at regular intervals of 20 pixels.  The grayscale stretch shows
the lowest light levels, permitting visibility of the large-separation
artificial sources while excluding the sources close to the star.  The
image measures approximately 40\asc\ $\times$ 10\farcs~~The
sensitivity curve derived from this image is the lowest one in Fig.\
\ref{fig:sensitz}.\label{fig:sensimg}}
\end{figure}

We have measured the detection limits for each star in the
survey as a function of angular separation from the central star.  To
do this we took the images of each star with 1\farcs0 seeing or
better---the acquisition of such images was a survey requirement (\S
\ref{chap4:obs})---and inserted artificial point sources at an array
of separations from the star.  These artificial point sources were
generated with appropriate Poisson noise statistics and angular sizes
to match the seeing conditions.  The magnitudes of these artificial
stars (calibrated to the photometric standards used during the
relevant observing run) were set so that each artificial star was just
visible to the eye.  From these magnitudes, the sensitivity curve is
derived.  An example of this is shown in Fig.\ \ref{fig:sensimg}.

\begin{figure}
\plotone{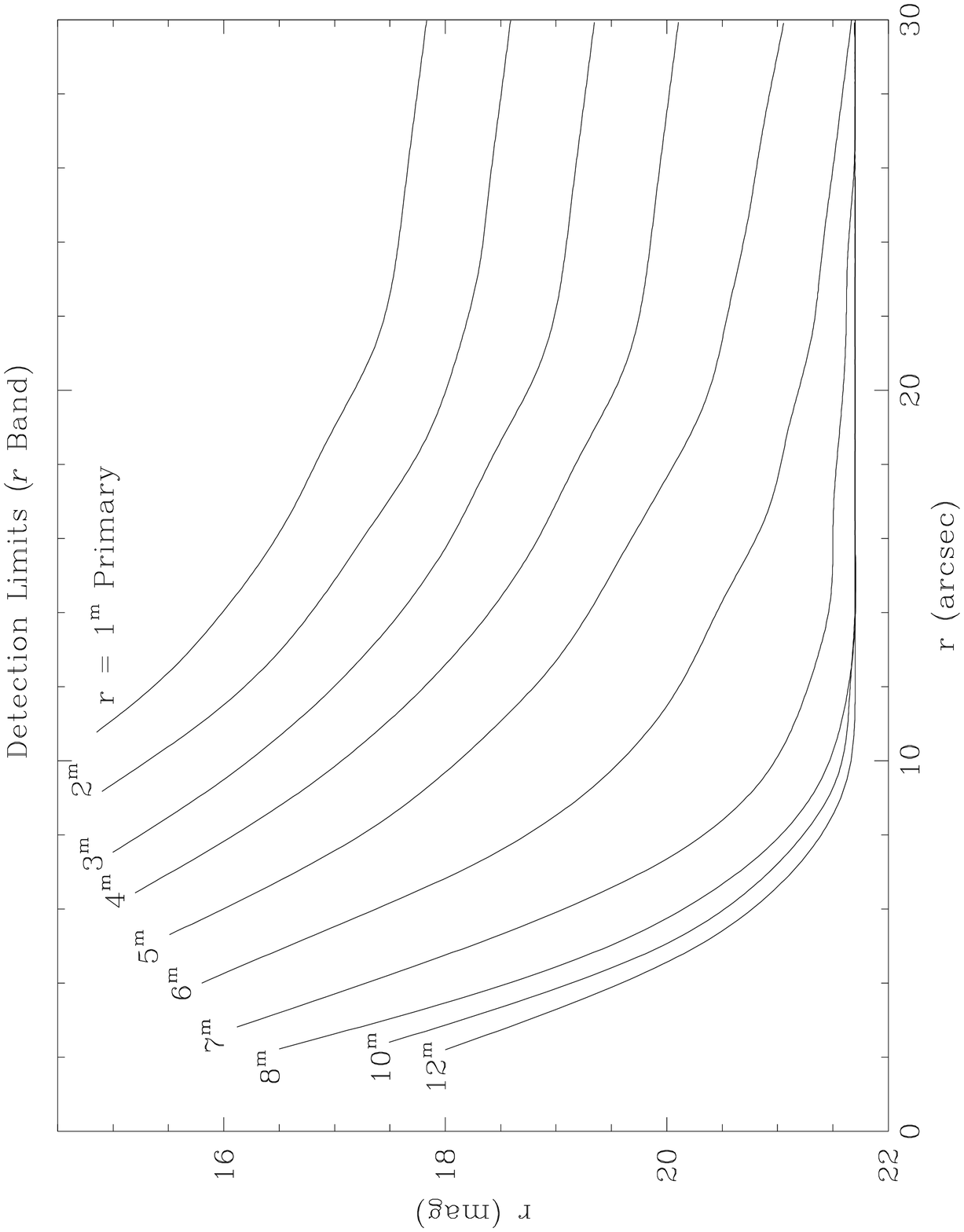}
\caption{Sensitivity curves for $r$
band showing magnitude limit as a function of separation (in
arcseconds) from the star for star magnitudes ranging from 1$^m$ to
12$^m$.\label{fig:sensitr}}
\end{figure}

\begin{figure}
\plotone{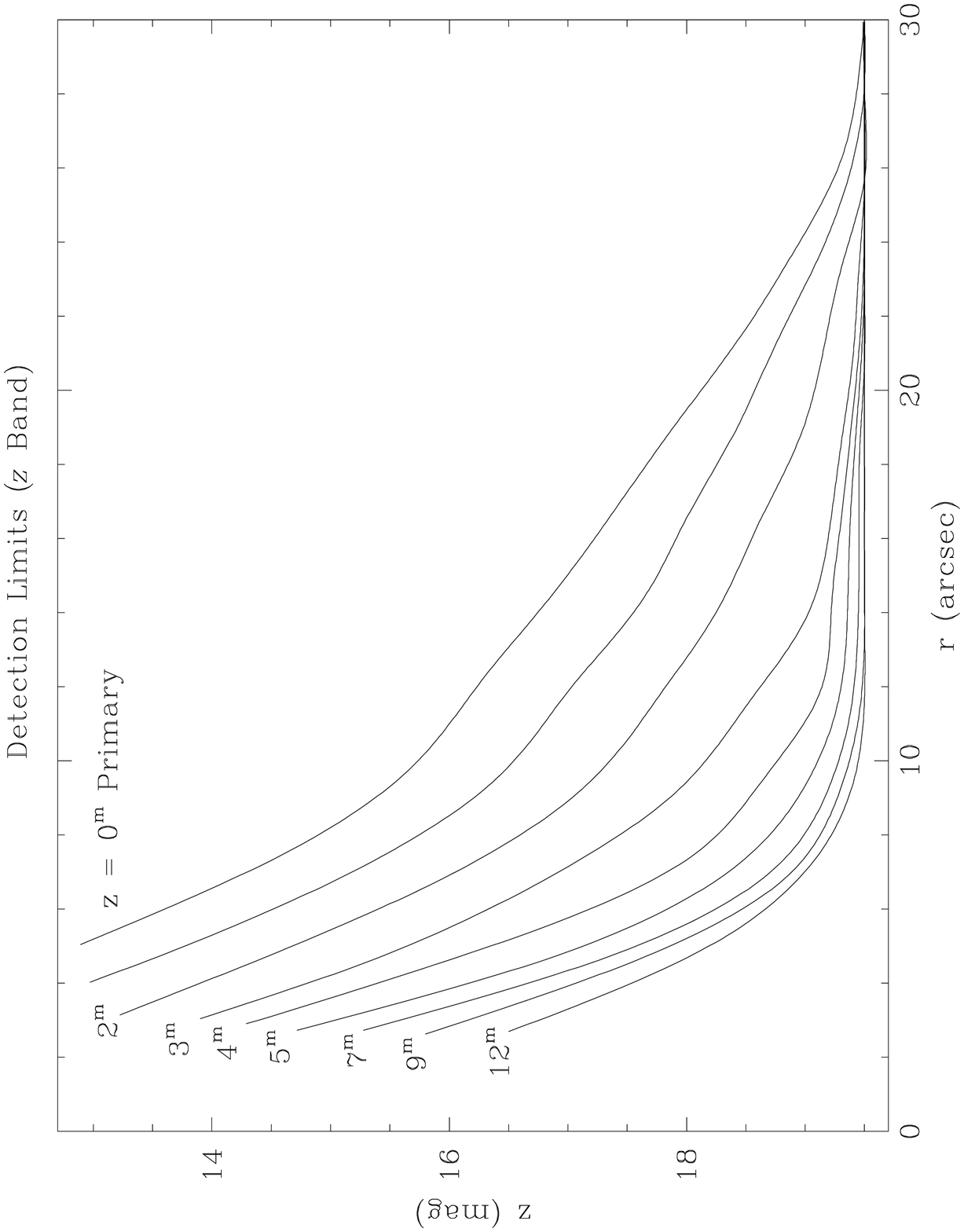}
\caption{Sensitivity curves for
$z$ band showing magnitude limit as a function of separation (in
arcseconds) from the star for star magnitudes ranging from 0$^m$ to
12$^m$.\label{fig:sensitz}}
\end{figure}

\begin{figure}
\plotone{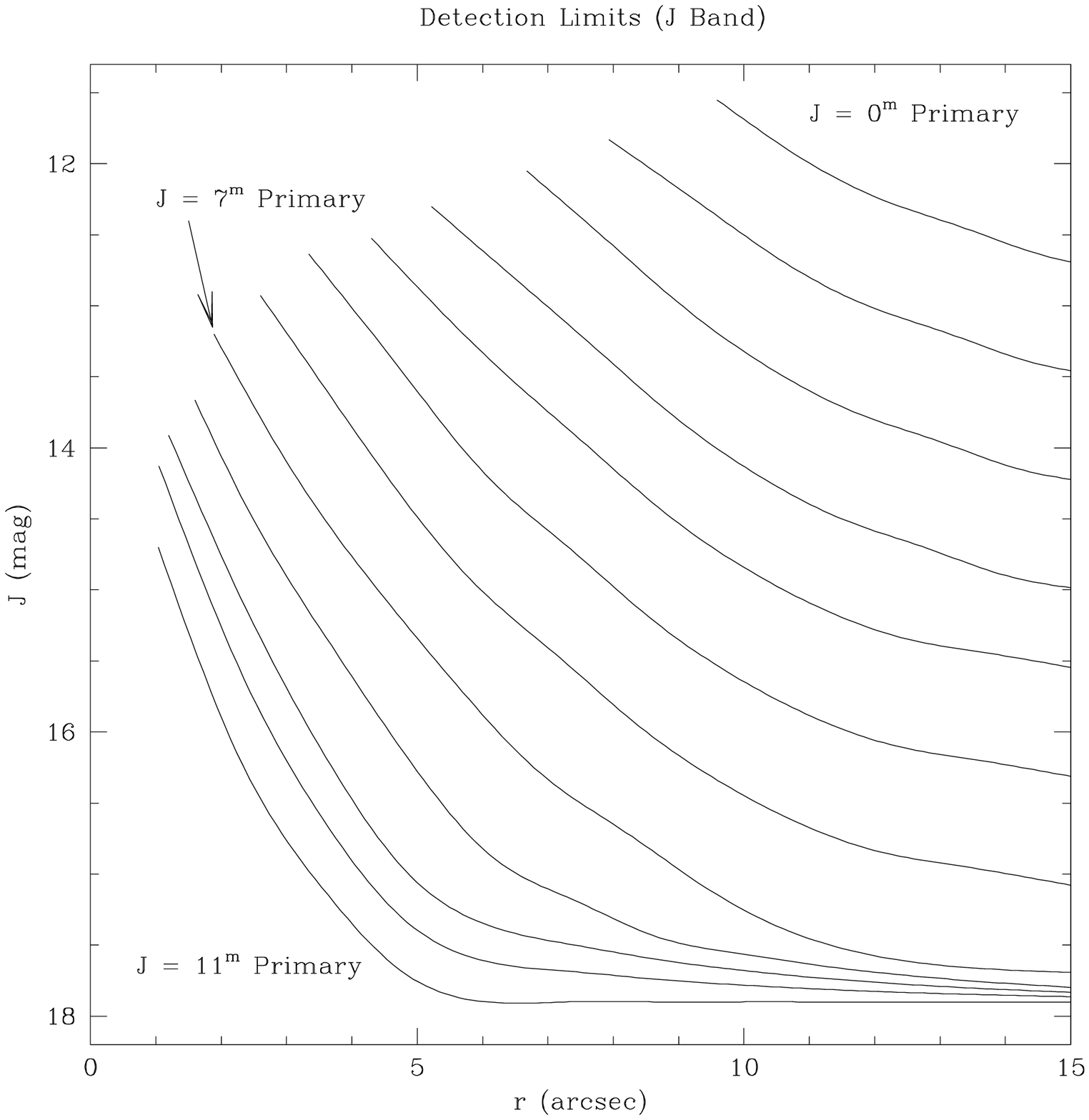}
\caption{Sensitivity curves for J
band showing magnitude limit as a function of separation (in
arcseconds) from the star for star magnitudes ranging from 0$^m$ to
11$^m$.\label{fig:sensitj}}
\end{figure}

\begin{figure}
\plottwo{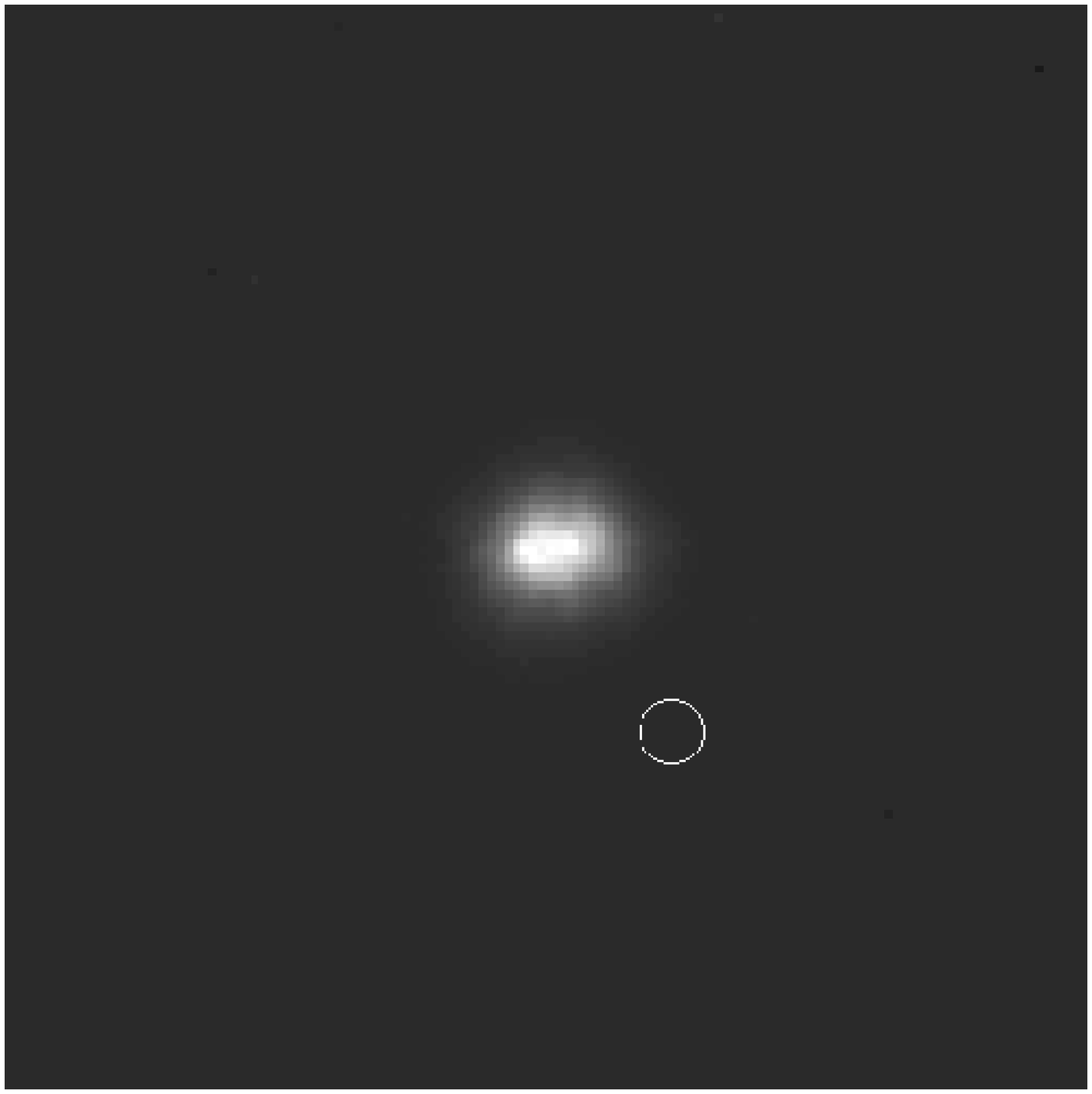}{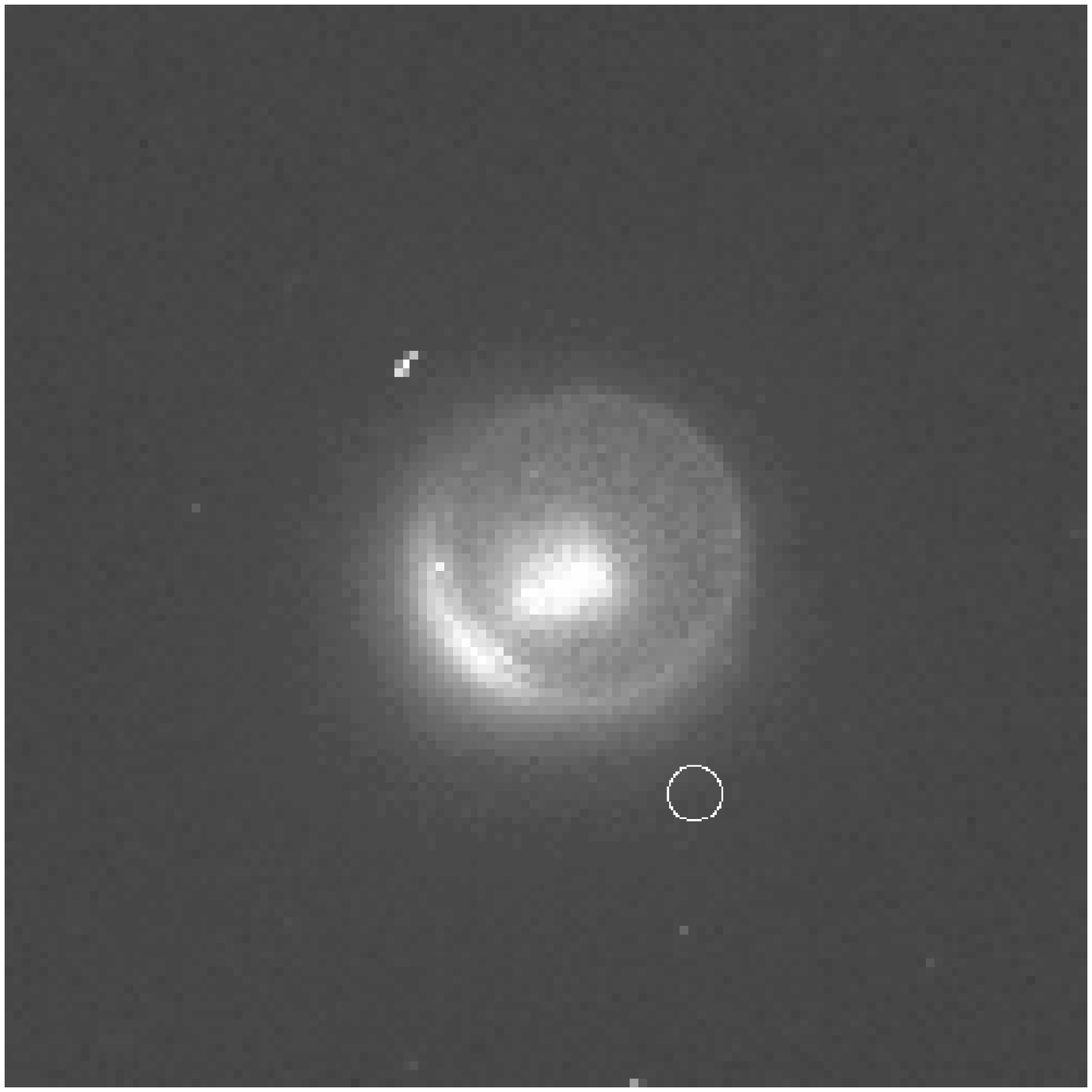}
\caption{Images of Giclas 089$-$032 (162.00).
This star has been resolved into two components.  The left image is a
5 s K band image taken in December 1996.  The components have a
separation of 0\farcs75, and North is up with East left.  The right
panel is a 1000 s $z$ band image taken in January 1998.  North is
4.2$^\circ$ left of up and E 90$^\circ$ counterclockwise from there.
The two components have been resolved through the semi-transparent
coronagraphic mask (4\farcs3 in diameter).  Here the components have a
measured separation of 0\farcs71, consistent, within the 0\farcs05
error bars, with the infrared offset measured 1.1 years earlier.  The
star has a known proper motion of 0\farcs354 yr$^{-1}$.  Thus this is
a common proper motion pair.  The circle in each panel indicates the
size of the core of a single star's point spread function.\label{fig:089}}
\end{figure}

From this procedure, we had a measurement of the faintest source
visible at a set of about 10 to 15 radii from the star.  Using a
spline interpolation between these points, we derived the magnitude
limit for $r$, $z$ and J as a function of radius measured from the
star.  In each band we have individual curves of this nature for each
central star in the survey.  These curves are summarized in Figs.\
\ref{fig:sensitr}, \ref{fig:sensitz} and \ref{fig:sensitj}, where we 
have displayed representative curves for various star brightnesses in
each bandpass.

There are several important effects documented by the curves in Figs.\
\ref{fig:sensitr}, \ref{fig:sensitz} and \ref{fig:sensitj}.  The most 
obvious is that the $z$ band imaging is the most sensitive, achieving
a maximum dynamic range of 15.5 magnitudes at 10\asc, while, in
addition, even the brightest companions can be imaged inside the
5\asc\ radius.  In comparison, the J band has no sensitivity at the
5\asc\ radius for the bright stars and only achieves a maximum dynamic
range of 13 magnitudes.  What is even more important is the large,
slowly-eroding wing of the point spread function in the J band.  The
$r$ and $z$ bands do not have nearly as much of this broad wing
primarily because of the pupil-plane stop in the coronagraph.  This
stop is designed specifically to depress the wings of the stellar
point spread function.

\section{Detection Limits for the Survey}

From the observational point of view, the curves in Figs.\
\ref{fig:sensitr}, \ref{fig:sensitz} and \ref{fig:sensitj} are the
ultimate measure of the sensitivity of our survey.  These curves
parameterize the sensitivity for every star in the sample.  However,
it is of paramount importance to convert the information in Figs.\
\ref{fig:sensitr}, \ref{fig:sensitz} and \ref{fig:sensitj} 
into statements about the survey as a whole and in terms of physical,
not observational, parameters.  For example, can we safely claim with
this data that for all the stars within 8 pc we would have detected
{\it any} unknown stellar companions in orbits between 3 and 200 AU?

To address this issue, we use the curves found in the previous section
to determine the number of surveyed stars for which we could have
detected a companion of a given magnitude at a given separation.  This
information is expressed in terms of the fraction (or percentage) of
the total number of stars imaged as a function of magnitude and
physical separation in AU.  To do this, we systematically went through
the catalog of stars.  For each star, we took the relevant sensitivity
curve, as derived in the previous section, and converted the magnitude
scale into absolute magnitudes, using the parallax of the star.  We
also converted the angular scale into physical separation in AU by
dividing by the parallax in arcseconds.  Then, we tested whether
companions with a set of magnitudes in each band and a set of
separations would be visible in our images.  By doing this exercise
for every star observed, we ended up with a tally of the number of
stars, as a function of magnitude and separation, for which the
analysis showed that a companion would be visible.

This analysis was done for each of the $r$, $z$ and J bands.  (The K
band sensitivity curves are essentially identical to the J band curves
so we did not independently conduct this analysis for K band.)  The
results are shown in Tables \ref{tab:Jmag}, \ref{tab:zmag} and
\ref{tab:rmag}.

The drop off in the survey sensitivity at large physical companion
separations is due to the varying physical field of view caused by the
distribution of parallaxes of the stars in the survey.  In the J band
there is no sensitivity outside of 120 A.U., for example, because the
field of view is 15\asc\ and the minimum parallax is 125 mas. 

In Tables \ref{tab:Jmag} through \ref{tab:rmag}, we have drawn a
horizontal line at the approximate absolute magnitude of a 0.08
M$_\odot$ star.  If the hydrogen burning mass limit is 0.08 M$_\odot$,
then all stellar companions must be brighter than the absolute
magnitude indicated by the line.  Brown dwarfs can be brighter than
this line if they are young, but objects below this line must be brown
dwarfs and not stars.  There are some indications that the hydrogen
burning mass limit may not be 0.08 M$_\odot$.  We refer the reader to
\citet{giz00} and references therein for observational evidence.

\section{Sensitivity to Stellar Companions}\label{chap4:senstar}

We now extend the analysis from the previous section.  Instead of
discussing the survey sensitivity in terms of observational
quantities, we relate those quantities to known properties of stars.
This allows us to determine the ability of the survey to find stellar
companions of the survey stars.  A stellar companion at the minimum
mass for hydrogen burning (0.08 M$_\odot$; Burrows et al.\ 1997) has
absolute magnitudes of $M_r = 17.4$, $M_z = 14.9$ and $M_J = 11.5$.
These magnitudes are determined by averaging the photometry of several
of the objects known to be at the minimum stellar mass \citep{HM93}.
(It is important to note, however, that the exact location of the
``hydrogen burning mass limit'' is not precisely known.  For example,
\citet{giz00} suggest that some L-dwarfs might be hydrogen burning
and yet have masses below 0.08 M$_\odot$.)
We now use these measurements to make a single table showing the
sensitivity to a minimum mass star in each of the bandpasses.  The
result is shown in Table \ref{tab:hbml}.

What this table demonstrates is that the combination of the infrared
and optical imaging permits the detection of any stellar companion at
separations greater than 10 AU.  We note that for the smaller
separations, J band is most sensitive.  $z$ band is more sensitive
than J band at the higher separations.  This is primarily due to the
drop in coverage at large separations in the J band (where the field
of view is only 32\asc).

Indeed, we have detected six new stellar companions of these stars.
These are described below.

\section{New Companions}\label{sec:comp}

In the course of our observations we have discovered or confirmed
seven new companions of nearby stars.  Three of these stars,
originally included in the 8 pc sample have been removed from the
sample because of new and more accurate trigonometric parallaxes.  The
new companions belong to the following systems: Gliese 105 (138.72),
Giclas 089$-$032 (162.00), Gliese 229 (173.19), Giclas 041-014
(224.00), LP 476$-$207, LP 771$-$095 (LTT 1445) and LHS 1885 (Giclas
250$-$031).  Only Gliese 229B is substellar.  The last three in the
list are no longer part of the 8 pc sample, which means that four of
our new companions are in the 8 pc sample.  Since Gliese 105AC (Fig.\
\ref{fig:c4f1}) and Gliese 229AB have been reported and described in
detail elsewhere, we simply refer the reader to Golimowski et al.\
(1995) for Gliese 105AC and Nakajima et al.\ (1995),
Oppenheimer et al.\ (1995), Matthews et al.\ (1996), Golimowski et al.\
(1998) and Oppenheimer et al.\ (1998) for Gliese 229AB.  Below we
describe the other companions.

\subsection{Giclas 089$-$032 (162.00)}

This star, with no trigonometric parallax measurement, is listed in
the CNS3 with a photometric parallax of 162.00 mas and a spectral type
of M 5.  The Palomar survey has resolved the star into a binary of
equal magnitude.  It was noted as a double source with 0\farcs7
separation in \citet{hen97}, but they had no information to
determine whether the two components were physically associated.  In
our coronagraphic images taken in January 1998, we resolve the two
components under the semi-transparent focal plane mask.  The short
infrared images also barely resolve the components.  With images taken
between December 1995 and January 1998 and the known proper motion of
the star, 0\farcs354 yr$^{-1}$, we have ascertained that the two
components exhibit the same proper motion and no measurable change in
relative offset during this time span.  Fig.\ \ref{fig:089} shows
several of our images.  Our measured separation is 0\farcs73.

\subsection{Giclas 041$-$014 (224.00)}

\begin{figure}
\plottwo{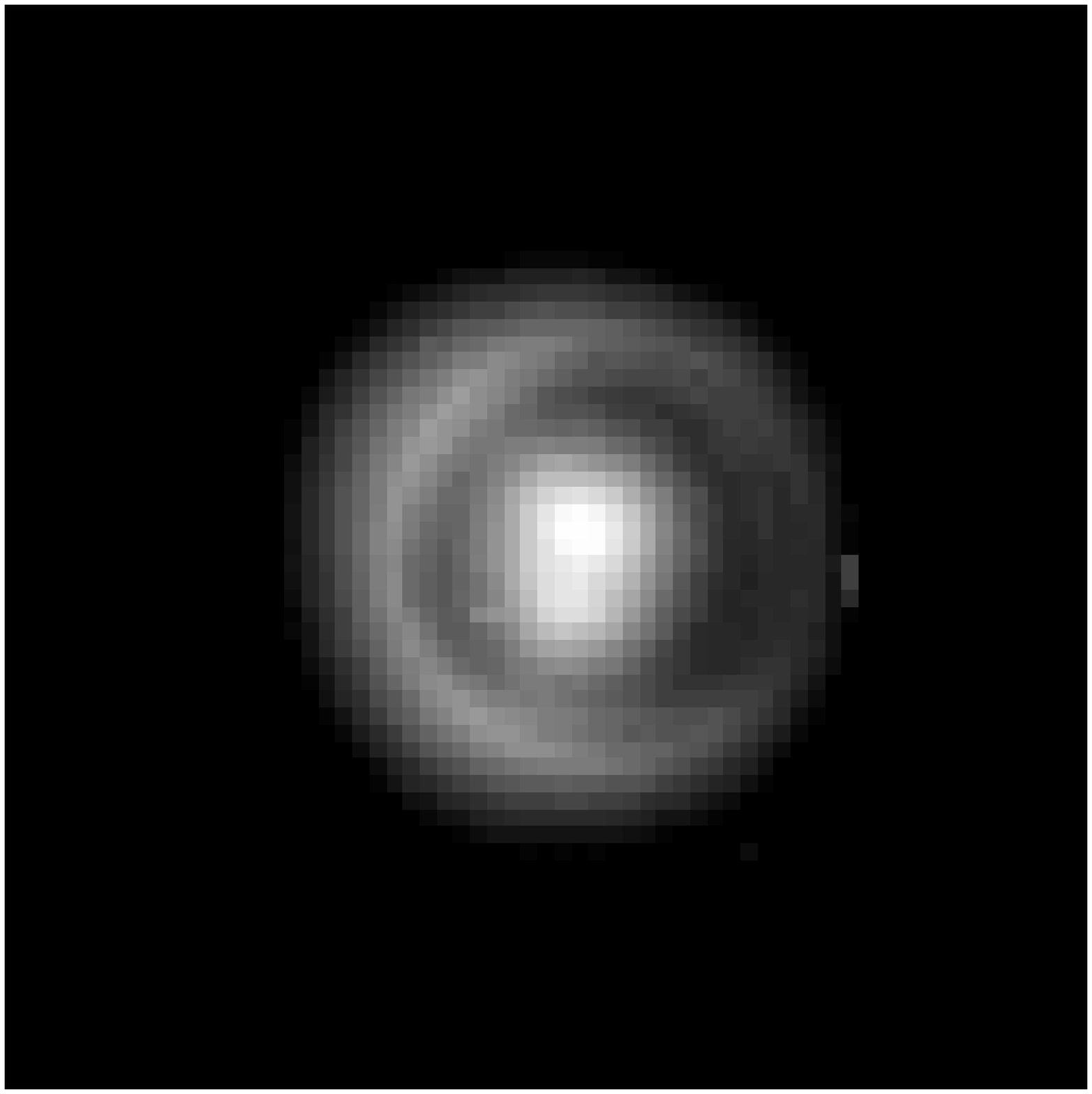}{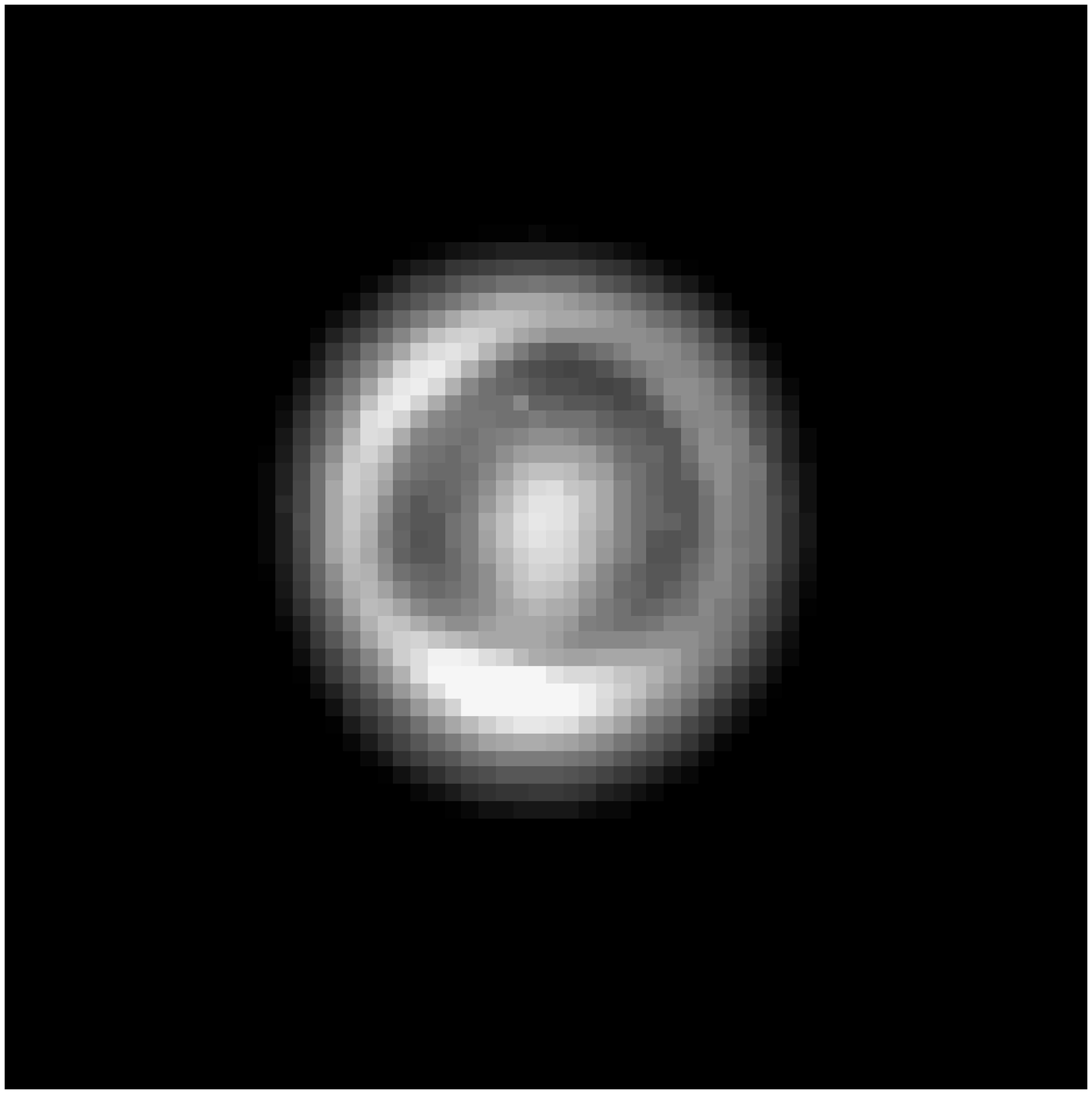}
\caption{Images of Giclas 041$-$014.
This star has been resolved into two components.  The left image is a
1000 s $z$ band image taken in November 1996.  The components have a
separation of 0\farcs47 $\pm$ 0.05.  North is left and East up.  The
right panel is a 1000 s $r$ band image taken in March 1998.
North is left and East is up.  The two components have been only
marginally resolved here through the semi-transparent coronagraphic
mask (4\farcs3 in diameter), the edge of which is visible because of
the spillover of light from the stars.  Here the components have a
measured separation of 0\farcs52 $\pm$ 0.1, consistent, within the
error bars, with the offset measured 1.25 years earlier.  The star has
a known proper motion of 0\farcs459 yr$^{-1}$.  Thus this is a common
proper motion pair.\label{fig:1405}}
\end{figure}

Giclas 041$-$014 is a star with a photometric parallax of 224 mas.
There is no trigonometric parallax measurement for this star.  Reid
and Gizis (1997) report that this object has a spectroscopic companion
of approximately equal mass.  Delfosse et al.\ (1999) have determined
the orbit of this companion (with a period of 7.6 days).  However,
Delfosse et al.\ also claim to have resolved a third component of the
system with adaptive optics images.  They did not publish a
confirmation of common proper motion for this object.  We have
determined that it is a physical companion.  They determine a
separation of 0\farcs62 and a difference of 0.5 magnitudes at K band.
This star, which is listed in Table 4.1 as a binary, was included in
our survey.  We observed it eight times over the duration of this
project. Only two of our observations were capable of resolving this
putative companion.  Both observations were with the AOC in extremely
good seeing conditions where the corrected image size was 0\farcs45
and 0\farcs50.  We resolved the companion and measured offsets of
0\farcs47 in November 1996 and marginally resolved the companion at
0\farcs52 in March 1998.  (The standard errors discussed in \S
\ref{sec:astrom} apply to the November 1996 measurement.  However,
since we only marginally resolved the two components in March 1998, we
suggest that the error on that measurement is $\pm$ 0\farcs1.)
Despite the marginal resolution in March 1998, the expected change in
relative offset of this star over this period of time is about
1\farcs~~Thus, if it were a background object, we would have easily
measured this large change in the offset.  The magnitude difference in
$z$ band is approximately 1.6.  Fig.\
\ref{fig:1405} shows the November 1996 $z$ and the March 1998 $r$ band
images.  In the case of this star (all three components of which are
included in Table 1), there was also a faint field star about 9\asc\
to the NW.  This star's relative offset between these epochs changed
approximately 1\asc, consistent with the 0\farcs459 yr$^{-1}$ proper
motion. 

\subsection{LP 476$-$207}

\begin{figure}
\plottwo{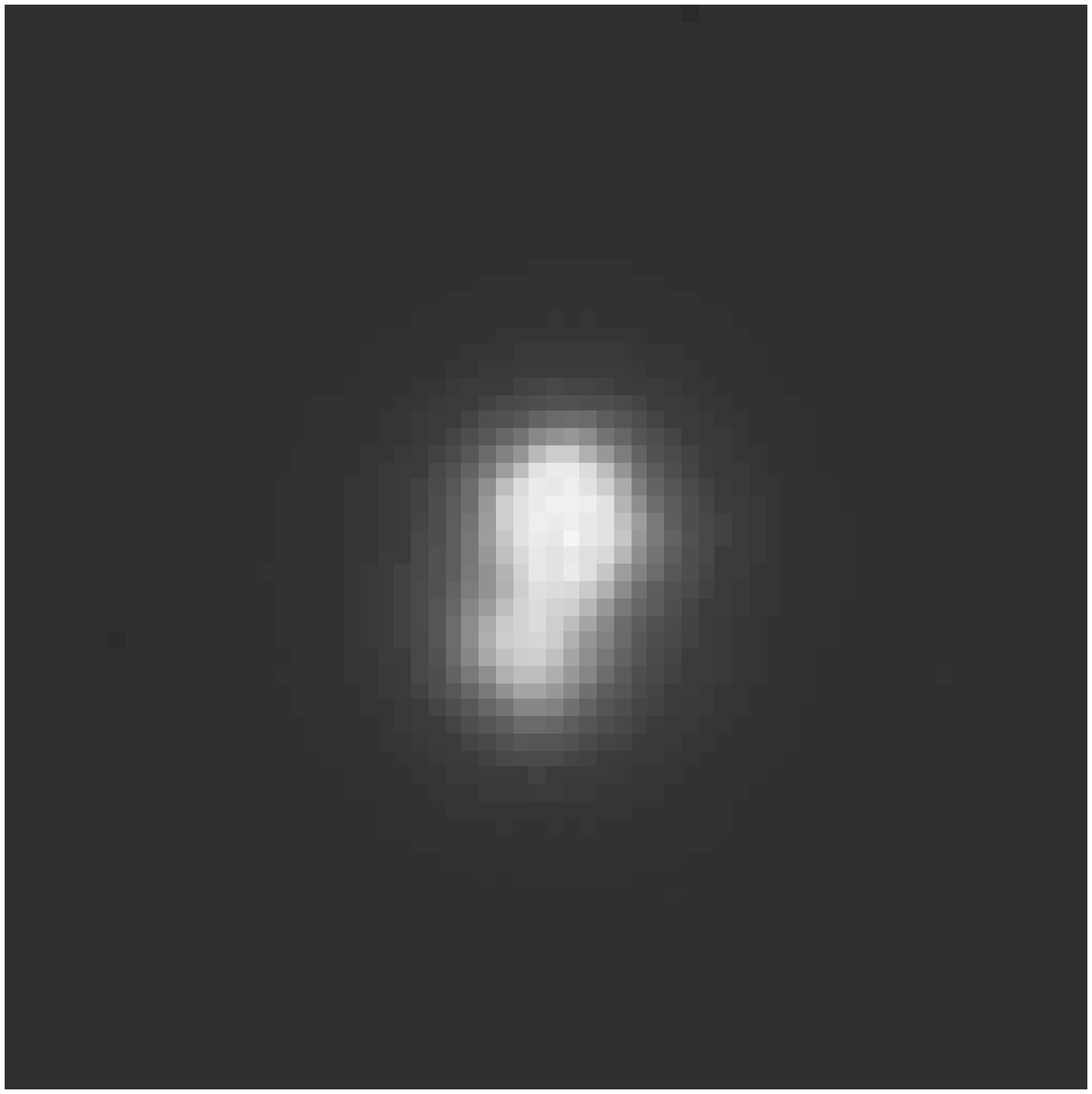}{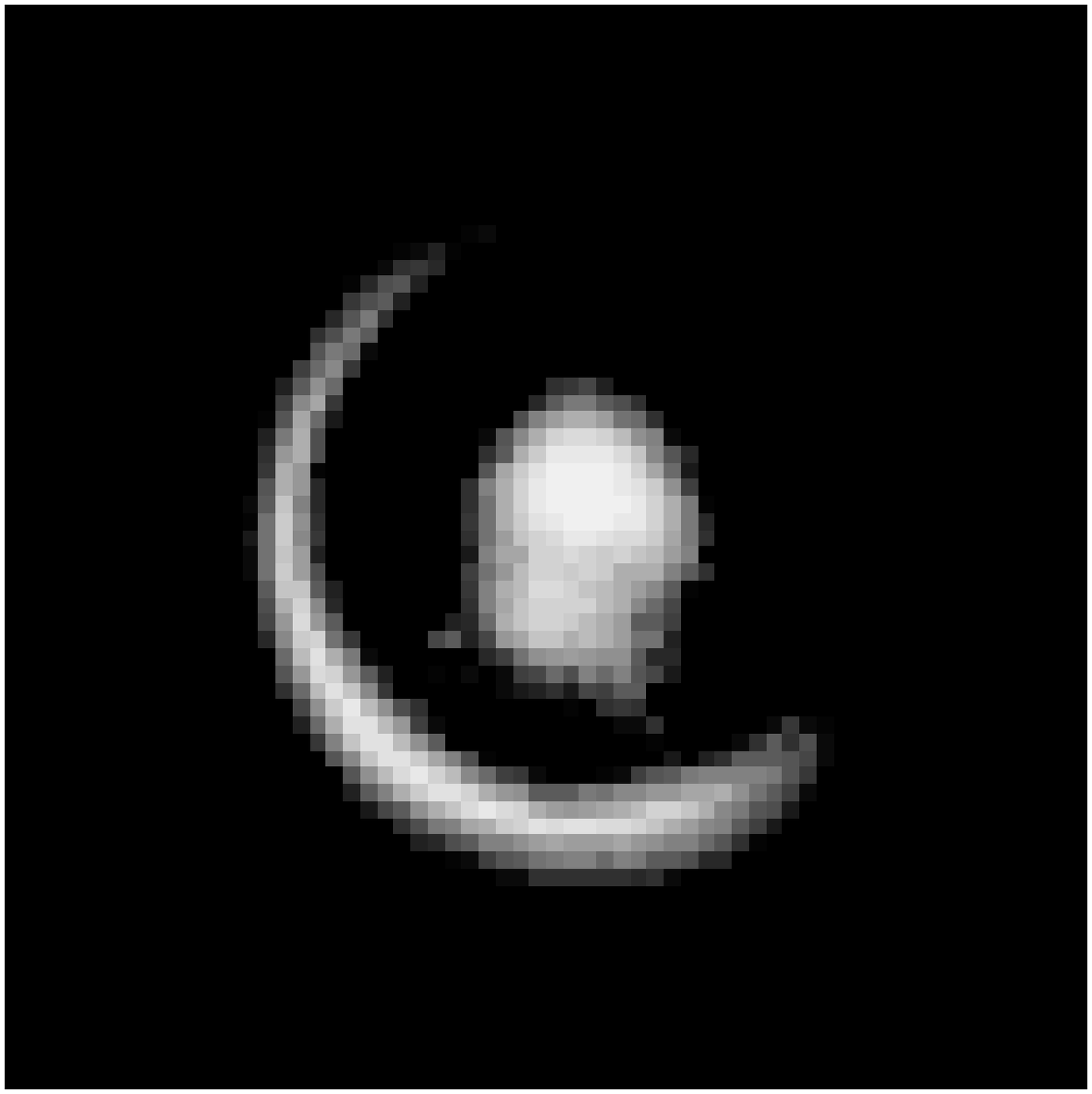}
\caption{Images of LP 476$-$207.
This star has been resolved into two components.  The left image is a
5 s K band image taken in October 1996.  The components have a
separation of 1\farcs03.  North is up and East is left.  The right
panel is a 1000 s $z$ band image taken in January 1998.  North is
up and East is left.  The two components have been resolved
through the semi-transparent coronagraphic mask (4\farcs3 in
diameter), the edge of which is visible because of the spillover of
light from the companion.  Here the components have a measured
separation of 0\farcs99, consistent, within the 0\farcs05 error bars,
with the infrared offset measured 2.24 years earlier.  The star has a known
proper motion of 0\farcs0837 yr$^{-1}$.  Thus this is a common proper
motion pair.\label{fig:476}}
\end{figure}

This M 4 star was given a photometric parallax of 142 mas in CNS3.
The subsequent Hipparcos measurement of 31.20 mas places it well
outside the 8 pc sample.  Part of the reason that the photometric
parallax is so incorrect must be due to the presence of the companion
we have found.  Even though this star is no longer in the 8 pc sample,
it was in our original catalog, so we observed it.  We found a common
proper motion companion about 1 magnitude fainter in K band than the
primary star.  This companion is located 1\farcs03 from LP 476$-$207.
The two images in Fig.\ \ref{fig:476} show a 5 s K band image from
October 1996 and a 1000 s $z$ band image from January 1998.  The
proper motion of this star is only 0\farcs0837 yr$^{-1}$, which is
less than 1 pixel yr$^{-1}$ in these images, but the 2.24 yr baseline
permits easy identification of this fainter object as a common proper
motion companion.  Henry et al.\ (1997) also identified this star as
double, without further information to determine that the two are
physically associated.  Delfosse et al.\ (1999) have confirmed the
results above, measuring an offset of 0\farcs97 and a magnitude
difference of 0.9 in the K band.  However, because of the single epoch
nature of their observation, they were unable to state with certainty
that this was a physical companion of the star.  Interestingly, they
also reported the detection of an unresolved spectroscopic companion
of the primary star.

\subsection{LP 771$-$095 (LTT 1445)}

\begin{figure}
\plottwo{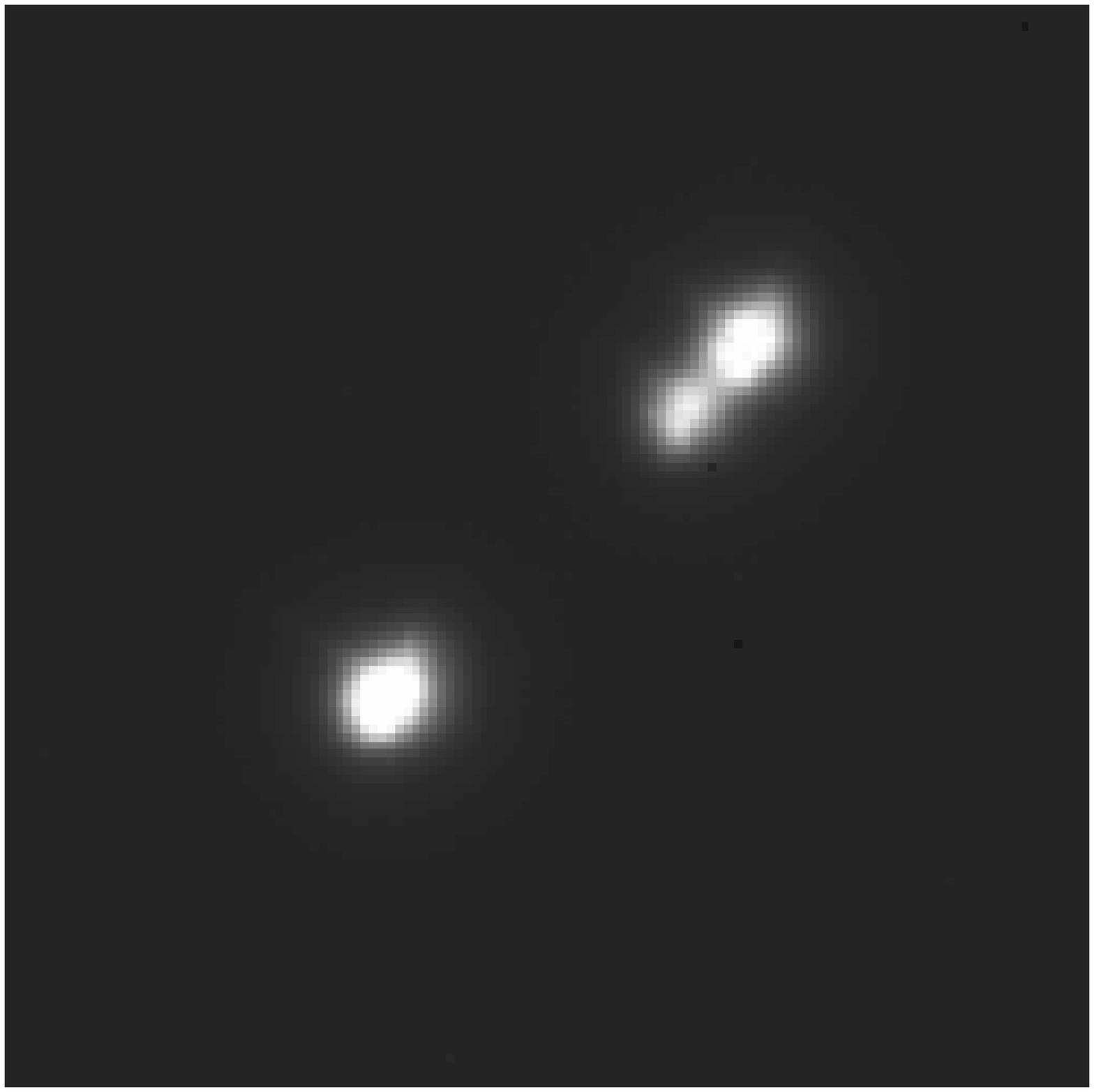}{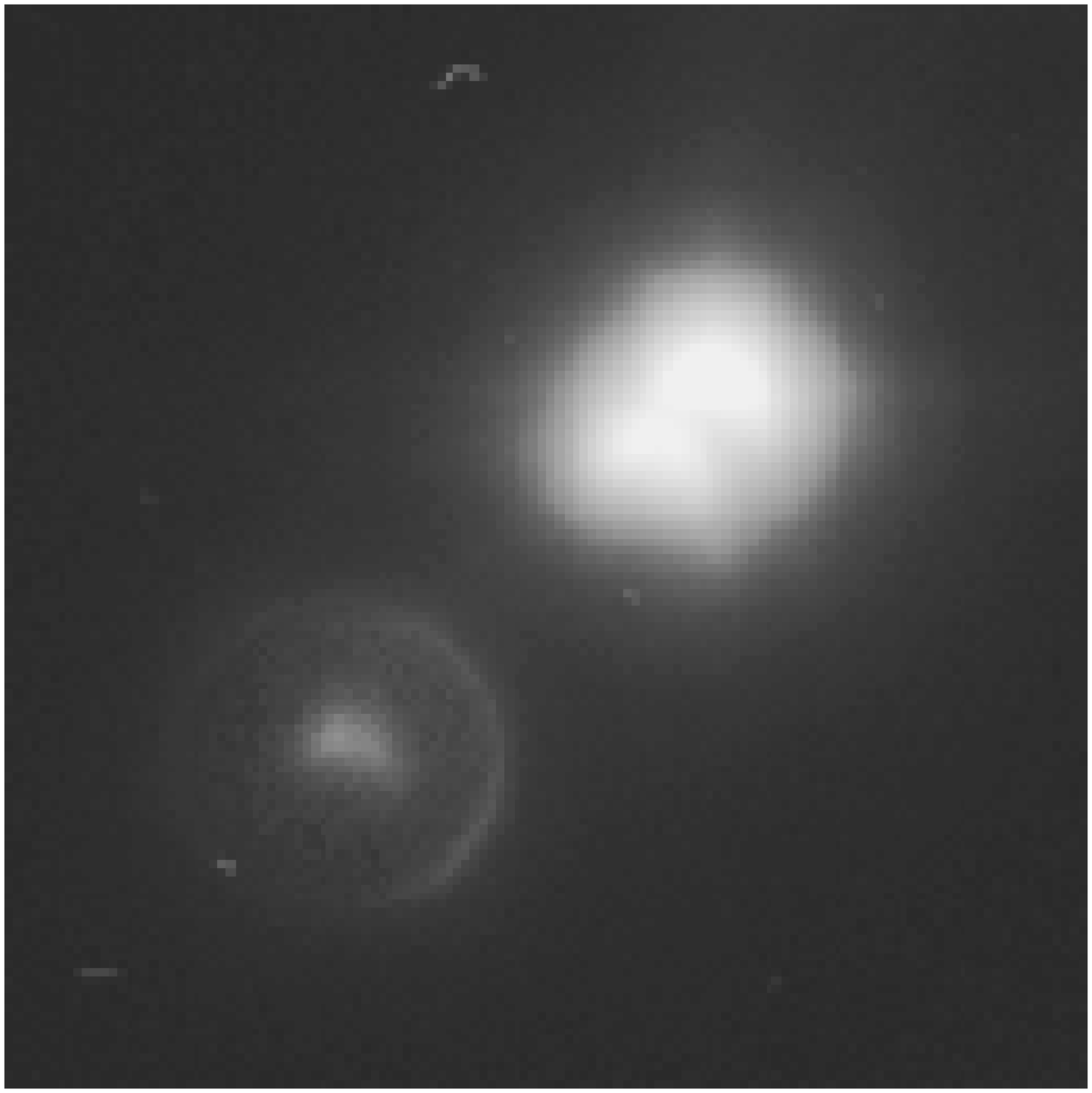}
\caption{Images of LP 771$-$095 triple system.
This star is a known binary (the two outer stars), but we have found a
third stellar component of the system which shares the proper motion
of the other two (LP 771$-$095 and LP 771$-$096).  On the left is a 5
s K band image taken in October 1996.  The components have a
separation of 1\farcs12 and 7\farcs23 from the star LP 771$-$095,
which is to the upper right in the images.  North is up and East
is left.  The right panel is a 1000 s $z$ band image taken in
October 1995.  North is up and East to the left.  LP 771$-$096 is
under the mask in this case.  All three components are clearly
visible.  Unfortunately, the astigmatism of the Palomar 60'' telescope
is also apparent.  In order to measure accurate astrometry on this
image, we used only the light in the brightest part of the point
spread function.  Our astrometry matches that from the infrared
images.  The proper motion of the outer pair of stars has been
measured to be 0\farcs4723 yr$^{-1}$.  Thus this is a common proper
motion triple system.\label{fig:771}}
\end{figure}

This star, along with LP 771$-$096, is a known binary, but we have
found a third component which sits along the line between the two
stars and which shares the proper motion of the binary.  The CNS3
listed a photometric parallax of 131 mas, but the Hipparcos mission
subsequently measured the trigonometric parallax at 92.97 mas.  The
proper motion of this star is 0\farcs4723 yr$^{-1}$, which makes
identification of common proper motion companions easy within a single
year.  The stars LP 771$-$95 and 771$-$96 were both classified as M
3.5 by \citet{JGthesis}, and the third component is approximately 1.2
magnitudes fainter than LP 771-095 in the K band.  It has a separation
of 1\farcs12 from LP 771$-$095.  LP 771$-$096 is 7\farcs23 from LP
771$-$095.  Fig.\ \ref{fig:771} shows two of the images we acquired of
this system.  
 
\subsection{LHS 1885 (Giclas 250$-$031)}

\begin{figure}
\plottwo{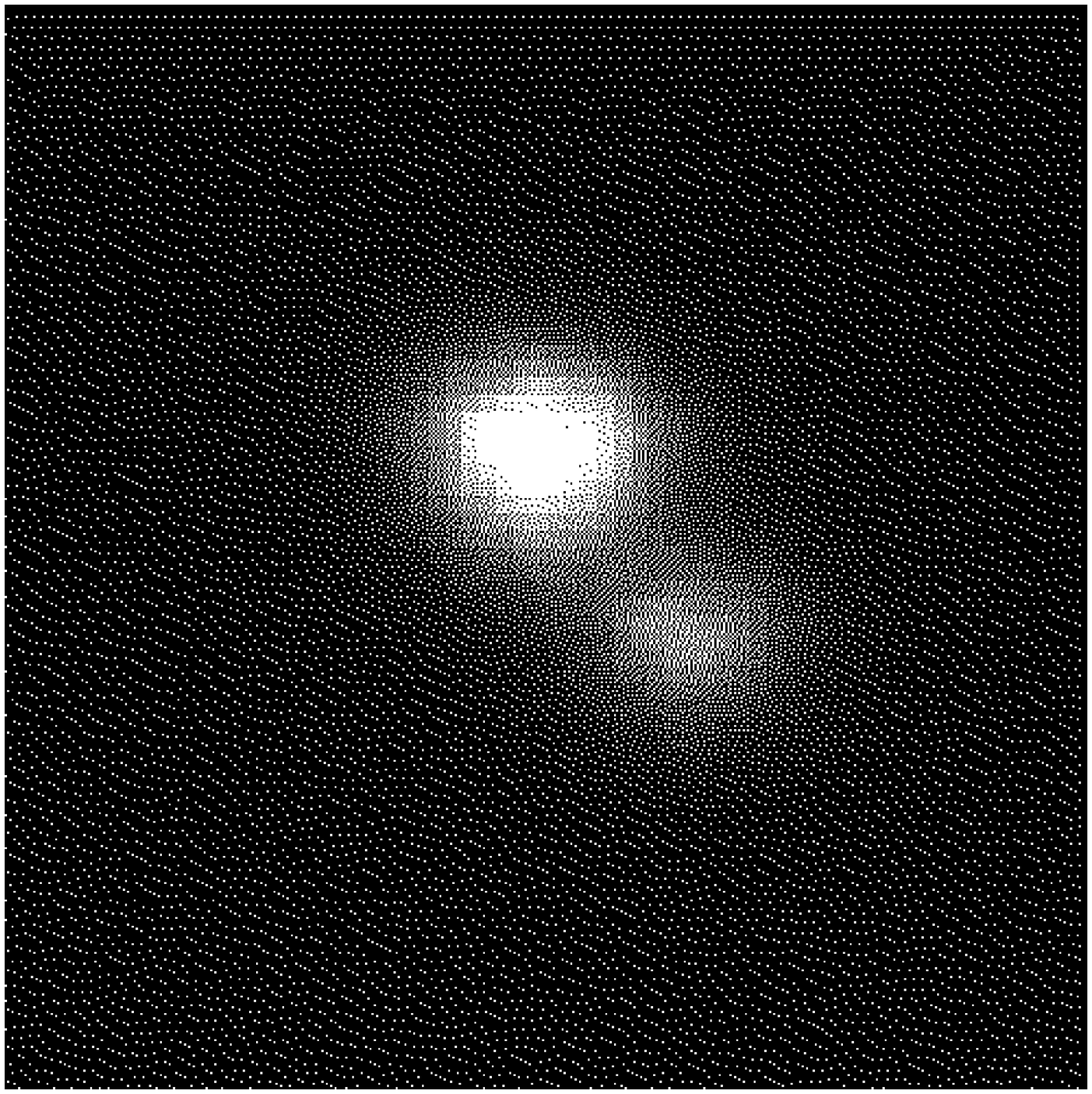}{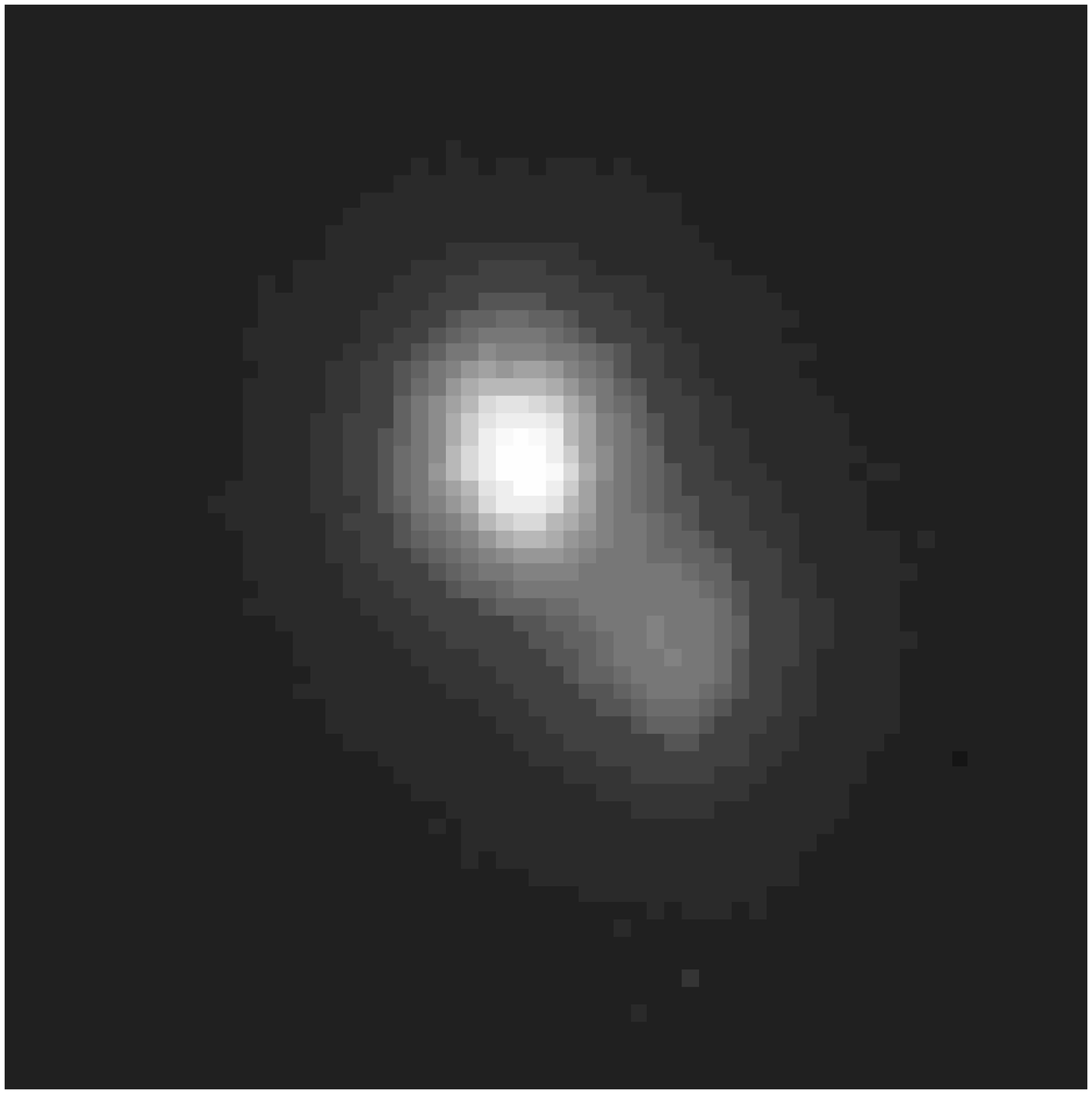}
\caption{Images of LHS 1885.  The second, 
fainter point source in these images is a common proper motion
companion of LHS 1885, the brighter star.  On the left is a 5 s K band
image taken in November 1995.  The components have a separation of
1\farcs65 and North is up with East left.  The right panel is a 5 s K
band image taken in December 1996.  North is up and East is left.  The
separation measured from this image is 1\farcs68.  The proper motion
of LHS 1885 is 0\farcs516 yr$^{-1}$.  Therefore, the fainter star is a
common proper motion companion of LHS 1885.
\label{fig:1885}}
\end{figure}

The CNS3 listed LHS 1885 with a photometric parallax of 129 mas.  The
Yale trigonometric parallax survey, however, measured a parallax of
87.4 mas, which removed it from our original sample of 8 pc stars.
This M 4.5 \citep{JGthesis} star was reported as double by Henry et al.\
(1997).  We have found that the second component shares the proper
motion of the star.  The second component is approximately 1.7
magnitudes fainter in the K band and is 1\farcs66 distant from the
primary star.  The proper motion of 0\farcs516 yr$^{-1}$ permits
identification of the common proper motion companion in less than a
year.  Fig.\ \ref{fig:1885} shows two K band images taken in November
1995 and December 1996.

\section{Sensitivity to Brown Dwarf Companions}

We now must expand the analysis from \S \ref{chap4:senstar} to fainter
levels and different colors: those of the brown dwarfs.  The use of
Gliese 229B photometry (Matthews et al.\ 1996) permits the production
of a table similar to Table \ref{tab:hbml}, but for cool brown dwarf
companions.  This is presented in Table \ref{tab:229B}.  We use Gliese
229B as a template cool brown dwarf because of all the cool brown
dwarfs, it is the most comprehensively studied \citep{O98}.  The only
other one with a known parallax is Gliese 570D \citep{bur00}, but its
photometry is not as comprehensively measured.

The contrast between Table \ref{tab:229B} and \ref{tab:hbml} is quite
dramatic.  The survey has no sensitivity to cool brown dwarfs in the
$r$ band.  This is principally because in this band the absolute
magnitude of the template brown dwarf, Gliese 229B, is 24.6
(Golimowski et al.\ 1998), while the images were limited to,
at best, 21.7.  The survey is most sensitive to brown dwarf companions
in the $z$ band at the wider separations and in the J band for
separations $\leq$ 40 AU.  This is due to the suppression of the
broad wings of the point spread function by the pupil-plane
coronagraphic stop.  Brown dwarfs in orbits with separations between
50 and 100 AU would have been detected around more than 80\% of the
stars in the survey.

\begin{figure}
\plotone{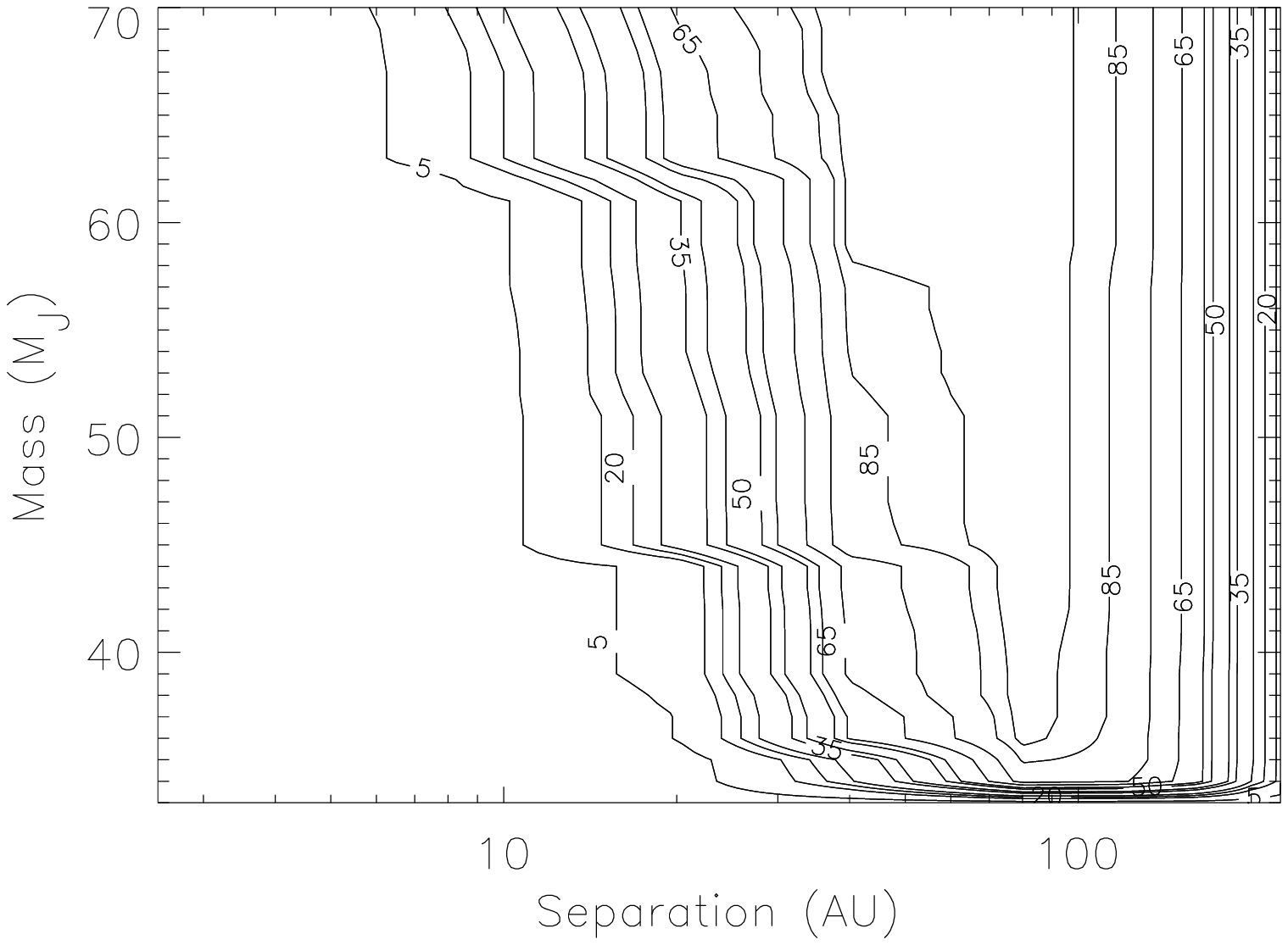}
\caption{Brown dwarf mass vs. separation: Survey coverage 
in $z$, age 5 Gyr.  This is a contour plot of an expanded form of the
data presented in Table \ref{tab:zmag5}.  It shows the survey
sensitivity in $z$ band as described in the text as a function of
brown dwarf mass and orbital separation, assuming the age of the stars
is 5 Gyr.
\label{fig:zmag5}}
\end{figure}

\begin{figure}
\plotone{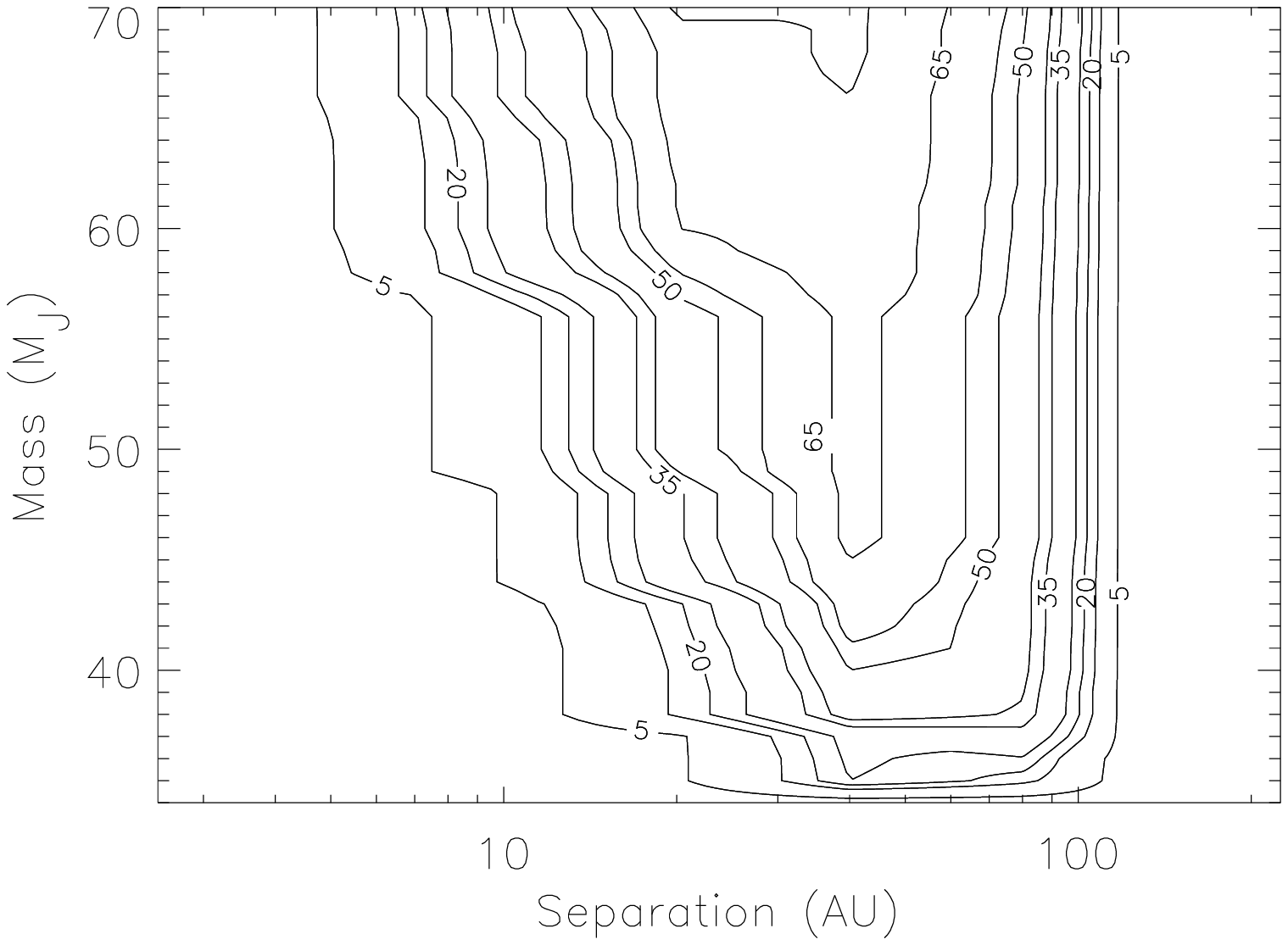}
\caption{Brown dwarf mass vs. separation: Survey coverage 
in J, age 5 Gyr.  This is a contour plot of an expanded form of the
data presented in Table \ref{tab:Jmag5}.  It shows the survey
sensitivity in J band as described in the text as a function of brown
dwarf mass and orbital separation, assuming the age of the stars is 5
Gyr.
\label{fig:Jmag5}}
\end{figure}

\begin{figure}
\plotone{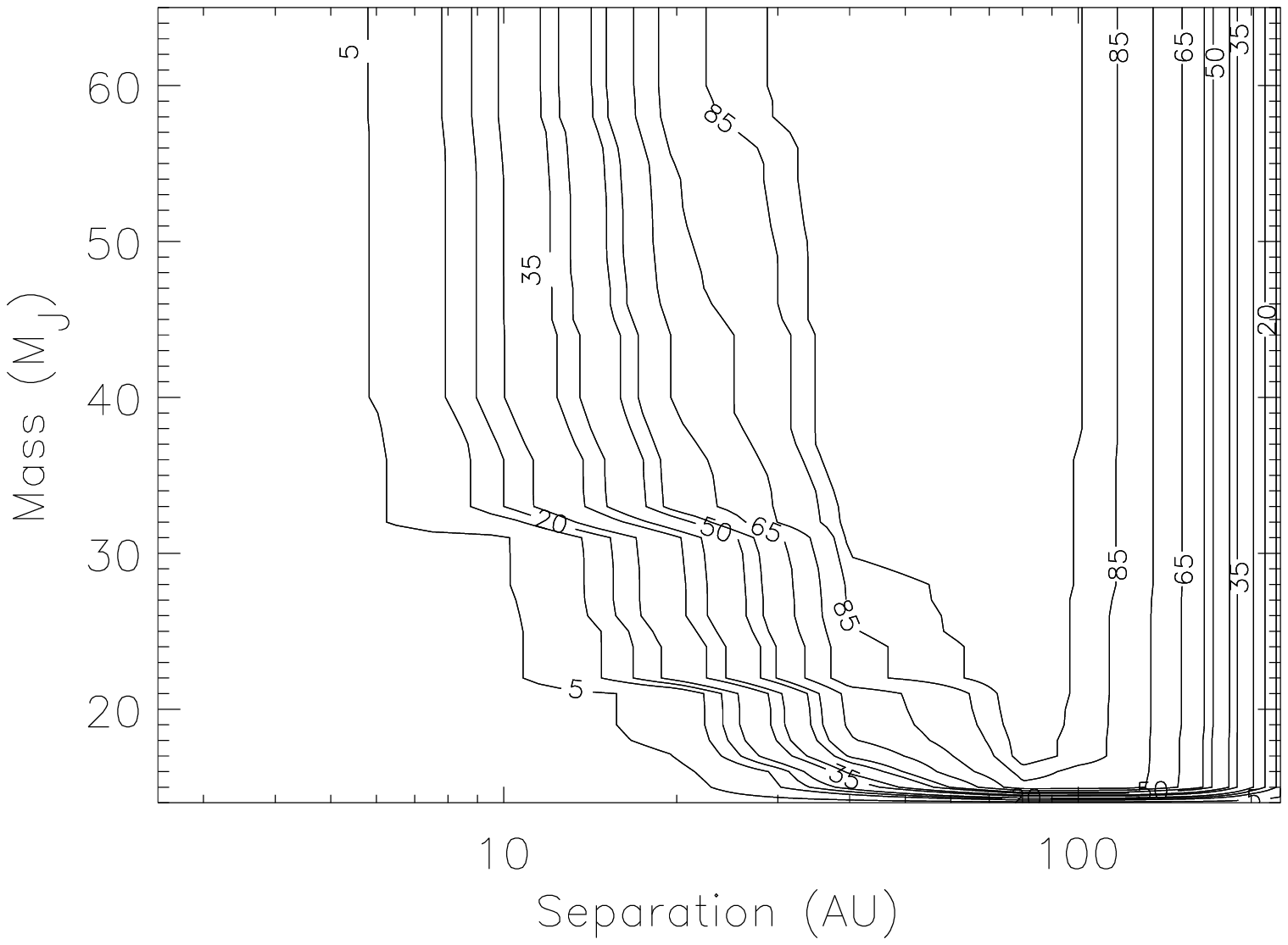}
\caption{Brown dwarf mass vs. separation: Survey coverage in 
$z$, age 1 Gyr.  This is a contour plot of an expanded form of the
data presented in Table \ref{tab:zmag1}.  It shows the survey
sensitivity in $z$ band as described in the text as a function of
brown dwarf mass and orbital separation, assuming the age of the stars
is 1 Gyr.
\label{fig:zmag1}}
\end{figure}

\begin{figure}
\plotone{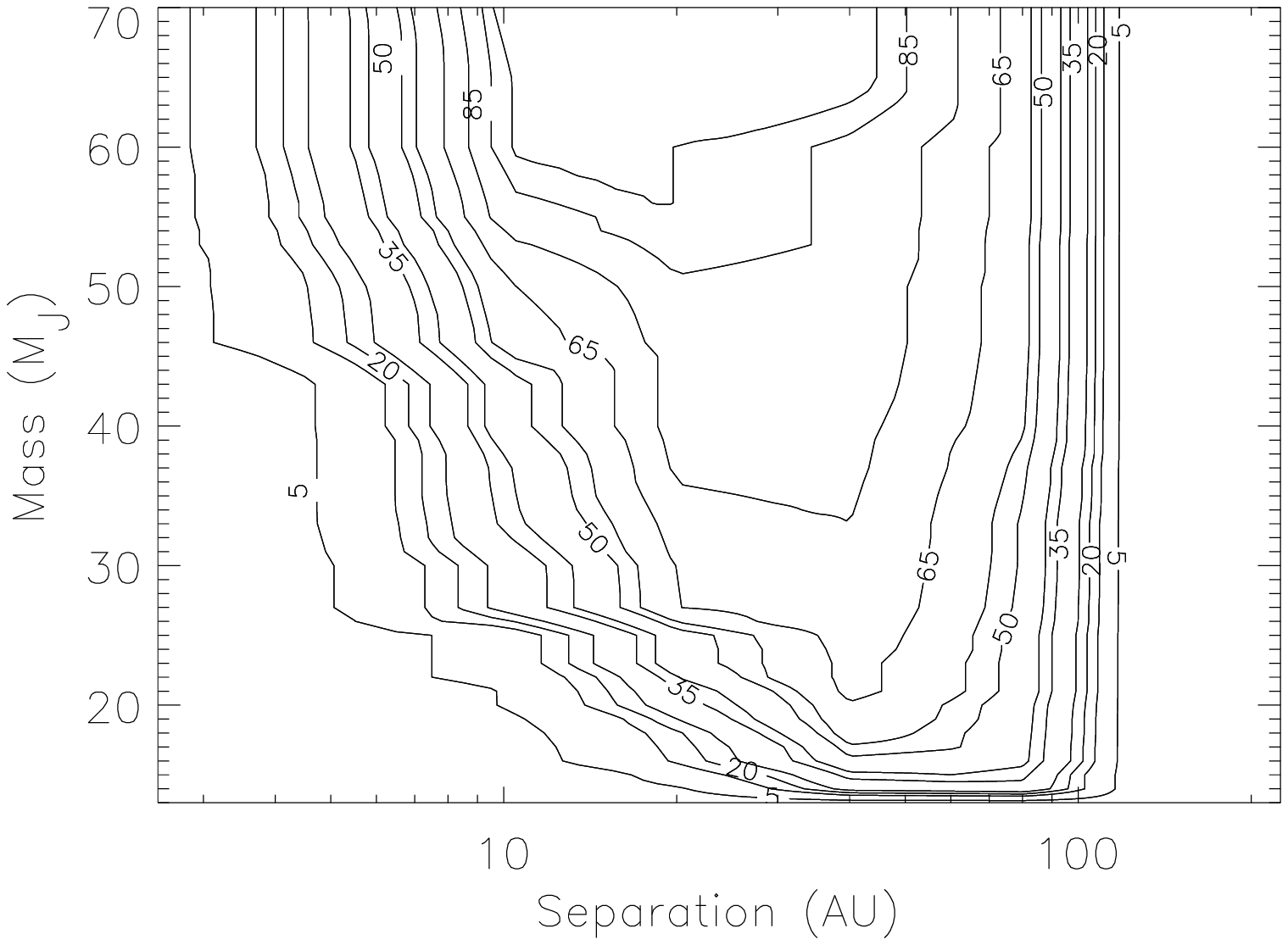}
\caption{Brown dwarf mass vs. separation: Survey coverage 
in J, age 1 Gyr.  This is a contour plot of an expanded form of the
data presented in Table \ref{tab:Jmag1}.  It shows the survey
sensitivity in J band as described in the text as a function of brown
dwarf mass and orbital separation, assuming the age of the stars is 1
Gyr.
\label{fig:Jmag1}}
\end{figure}

Ultimately we would like to express the sensitivity of the survey in
terms of what the lowest mass brown dwarf we could possibly detect is.
This is complicated by the fact that brown dwarfs cool.  As such, a
brown dwarf of a given mass will evolve through many magnitudes of
brightness in a given bandpass over the time scale of several billion
years.  Unfortunately, the state of the art models which produce
synthetic spectra, and thus color information, do not extend to the
$r$ band.  However, the models of Burrows et al.\ (1997, 2000;
personal communication) supply us with the magnitudes in the $z$ and J
bands for brown dwarfs of all masses (15 M$_J$ to 70 M$_J$) for two
different ages, 1 and 5 Gyr.  From this information we can convert the
absolute magnitudes in Tables \ref{tab:zmag} and \ref{tab:Jmag} into
brown dwarf masses for each of the two ages, 1 and 5 Gyr.  The results
are shown in tabular and graphical form: Tables \ref{tab:zmag5}
through \ref{tab:Jmag1} and Figs.\ \ref{fig:zmag5} through
\ref{fig:Jmag1}.

We should note that the two ages chosen here are unfortunately not
perfectly representative of the ages of the sample stars.  Assuming a
constant star formation rate in the galaxy, the ages of stars in the
disk would be evenly distributed between 0 and 10 Gyr.  Unfortunately
we do not have model flux densities for 10 Gyr objects.

\noindent{\bf Caveat}

We do not separately evaluate the conversion between J magnitudes and
mass and K band magnitudes and mass because the measurements in the
two bands yields essentially identical masses.  As we mentioned above
(\S \ref{chap4:senstar}), the sensitivity curves for K band are the
same as those for J band.

\section{Discussion and Summary}\label{sec:chap4res}

What is becoming increasingly clear from the searches for field brown
dwarfs, such as the Two-Micron All-Sky Survey \citep{reid99}, is that
brown dwarfs greatly outnumber stars in the field population.  It
would seem to follow logically that many stars could have brown dwarf
companions, if the formation mechanism for binary stars applies to
star-brown dwarf systems.  However, we have shown here that brown
dwarfs seem to have a multiplicity fraction with stars far below the
17 to 30\%\ observed for all stars \citep{reid97}. Between 40 and 100
AU, we would have detected brown dwarfs more massive than 40 M$_J$,
around 80\%\ of the survey stars.  The only other cool brown dwarf
companion of a star within 8 pc is Gliese 570D
\citep{bur00}.  We did not detect this object because the separation
is $\sim$ 4 arcmin, placing it outside our field of view.

The initial goal of this survey was to find brown dwarfs.  However,
because we found only one, and because our survey detection limits are
complex functions of brightness, separation and age, placing
constraints on possible mass and separation distributions of brown
dwarf companions requires extensive Monte Carlo simulations.  This is
a rather complex problem which requires its own computational
techniques.  This work is currently in progress and will be published
in a separate paper.

Here we present several simple statements which can be made with
certainty.

1. This survey would have found all stellar companions of any type
around 98\%\ of the survey stars and between 3 and 30\asc\ of the
stars.  Indeed in \S \ref{sec:comp} we present 6 new stellar
companions.  This does not dramatically change the multiplicity
fraction for the 8 pc sample.

2. Brown dwarfs more massive than 40 M$_J$, at least as old as 5 Gyr,
would have been detected around 80\% of the survey stars for
separations between 40 and 120 A.U.  Only one such object exists
(Gliese 229B, at $>$ 39 A.U.), implying a binary fraction of around
1\%, assuming that Gliese 229B is a prototypical brown dwarf.  (We
must note here that the exclusion of models of older brown dwarfs in
this assertion must be considered when interpreting the result.  A
proper assessment of the constraints provided by our survey on the
binary fraction of brown dwarfs really requires extensive modeling of
the possible populations of stars and brown dwarf companions.)

3. There has been no complete assessment of the population of brown
dwarf companions of the survey stars for separations outside 100 A.U.
The most complete study to date has been that of Simons et al.\
(1996), but it did not cover the whole sample and turned up no new
brown dwarfs.  Our survey is insensitive to such wide separation
binaries.  However the 2MASS project \citep{bur00} should reveal all
companions of the known nearby stars with wide separations that are
similar to or hotter than Gliese 229B.

4. A more sensitive survey of the same stars in the sample presented
here is necessary to obtain a complete census of brown dwarf
companions in the solar neighborhood.  This requires the
suppression of scattered light from the primary stars (i.e.\ achieving
a higher dynamic range) and an increase in the limiting magnitude of
the sky-limited regions of the images.  The next step in this sort of
research is a full-scale, adaptive-optics-based survey, ideally with
simultaneous infrared and optical imaging.

\subsection{Endnote}

In \S \ref{chap4:sens} we addressed the issue of how sensitive the
survey is as a whole and for individual stars.  Ideally these
calculations and observations should be gathered into a general
statement about brown dwarf companions of nearby stars.  This turns
out to be a rather complex problem with a large and essentially
unexplored parameter space.  It is unexplored not only from the
observational standpoint but essentially no research has been
conducted on the theoretical aspects of the problem.  

\begin{figure}
\plotone{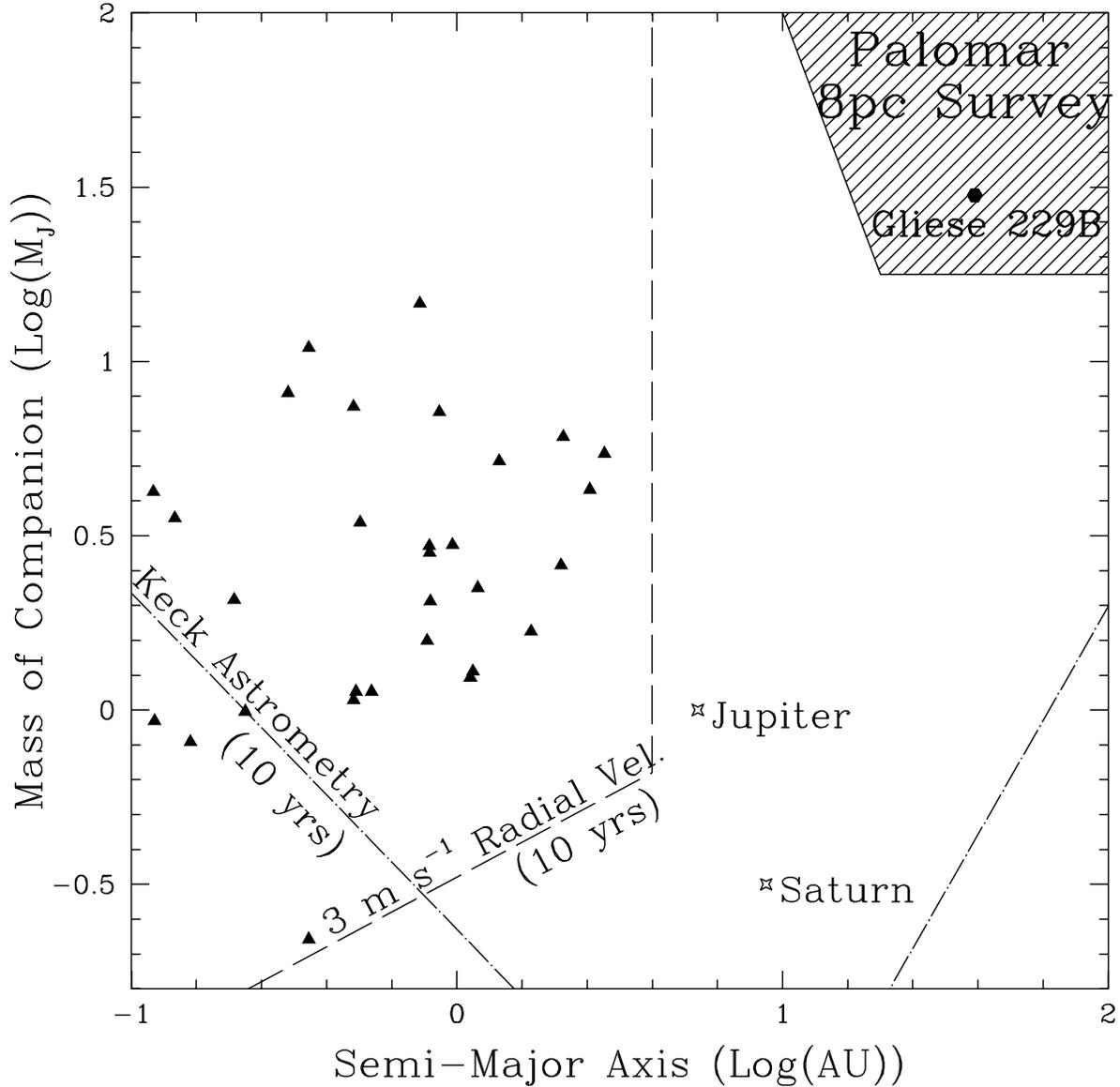}
\caption{Mass-separation parameter space.  
This plot shows the known substellar companions of stars discovered
to-date.  Jupiter and Saturn are indicated and labelled.  Gliese 229B
is represented by a black circle.  The planets found
in radial velocity searches are represented by closed triangles.  The
curves indicate the detection limits of several techniques.  The
dashed line shows the 3 m s$^{-1}$ limit of the current radial
velocity searches with baselines of 10 years.  The dash-dot lines show
the predicted limits of an astrometric search using the Keck
interferometer project over a 10 year period of observations.  Our
survey probed the shaded region in the upper right.  Our work
represents the first direct imaging project to probe this parameter
space.
\label{fig:radial}}
\end{figure}

From the observational standpoint, the search for faint or low-mass
companions of stars has only become practical in the past five to ten
years.  The two approaches to the problem---direct and indirect
detection---have turned up positive results, but each has access to a
different part of the parameter space.  The parameter space is defined
by mass of the companion and orbital separation.  This seems simple
enough, but as shown in the previous section the sensitivity of a
direct observing campaign is not a constant through any region of this
parameter space when the mass is below the ``hydrogen burning limit.''
This is particularly true because brown dwarfs cool.  The cooling
essentially introduces an additional parameter, the age.  In the case
of the indirect searches the sensitivity to mass is uninfluenced by
age, but the parameter space is also explored in a non-uniform manner
for a large sample of stars: it depends mainly upon the length of time
over which the observations are scattered for each star and how they
are distributed in time.  For example, periodic observations will be
completely insensitive to objects which orbit with a multiple of the
period of observation.  The point of this discussion is that the
mass-separation parameter space is poorly sampled, and making direct
comparisons between the direct and indirect observing methods is
difficult because the overlap in parameter space is only now beginning
to exist.  The simplest comparison is shown in Fig.\ \ref{fig:radial},
where the mass-separation parameter space is shown with lines
indicating sensitivity limits of various search techniques.  The only
complete imaging survey on the plot is that presented in this paper.

The only certain statement we can make at this time is that the
multiplicity fraction of brown dwarfs is far smaller than the 35 to
40\%\ for stellar binary systems.  In light of this it is important to
discuss the mass function.  The mass function below the ``hydrogen
burning limit'' has been the subject of heated debate and has mainly
relied upon observational rather than theoretical constraints.
(i.e. this part of the mass function cannot be calculated
theoretically at present.)  In the past ten years it has become clear
that the Salpeter mass function, which works for higher mass stars,
does not apply to the very lowest mass stars.  Recently, studies of
open star clusters such as the Pleiades, which probe into the brown
dwarf mass range, have begun to provide extensions of the mass
function.  (See, for example, Mart\'{\i}n et al.\ 1998.)  However, the
masses of the objects discovered are generally not well-constrained
because the theoretical models of these objects are not complete and
unable to reproduce all of the observations.  Other techniques for
finding brown dwarfs in the field, such as the microlensing
experiments, give accurate masses, but have found such a sparse number
of objects in the brown dwarf mass range that the error bars on the
implied mass distribution are large.  A careful analysis of the MACHO
results (Alcock et al.\ 1998) is presented by Chabrier and M\'{e}ra
(1998) and ``clearly illustrates the difficulty to reach robust
conclusions about the mass in the form of substellar objects in the
central regions of the Galaxy, and more precisely, in the disk and the
bulge, from present microlensing experiments.''  Indeed, Chabrier and
M\'{e}ra cannot constrain the space density of brown dwarfs to a range
smaller than an order of magnitude around 9 $\times$ 10$^{-3}$
M$_\odot$ pc$^{-3}$.

A further complication of this problem stems from the observation,
most recently by Reid and Gizis (1997) and Reid et al.\ (1999), that
the mass function for companions is actually different from the mass
function of field stars.  In their analysis of the 8 pc sample, they
find that the distribution of the mass ratios of multiple systems has
a significant peak near 0.95.  In their estimation this
excludes the notion that companions of stars come from the same mass
function as solitary stars.  For a survey of the nature presented
here, this makes drawing conclusions about the various mass functions
described above essentially irrelevant.  It would be akin to trying to
understand techno music by listening to classical violin concertos.
There has been no study of companion mass functions in the brown dwarf
regime.  Furthermore, if the mass ratio distribution that Reid and
Gizis (1997) find extends into the brown dwarf mass range, our survey
excludes the most important set of stars to find brown dwarf
companions of: we argued that the 25\%\ incompleteness of our sample
is all due to missing the very lowest mass stars within 8 pc.  These
are the ones that would be expected to have more brown dwarf
companions.

\clearpage

\clearpage

\clearpage

\clearpage

\clearpage

\clearpage

\clearpage

\clearpage

\clearpage

\clearpage

\clearpage

\clearpage

\clearpage

\clearpage

\clearpage

\clearpage

\clearpage

\clearpage

\clearpage

\clearpage

\clearpage

\begin{deluxetable}{lccrrrllccc}
\tablecolumns{11}
\rotate
\tabletypesize{\scriptsize}
\tablewidth{551pt}
\tablecaption{The 8 pc Sample\label{chap4:tab1}}
\tablehead{
\colhead{Parallax} & \multicolumn{2}{c}{Position (J2000.00)} &
\multicolumn{2}{c}{$\mu$ (mas)} & \colhead{V\phantom{i}} & \colhead{CNS3} & \colhead{Durch.} & \colhead{Other} & \colhead{SMA} & \colhead{Source}\\
\colhead{(mas)} & \colhead{RA ($^h$ $^m$ $^s$)} & \colhead{Dec. ($^\circ$ \arcmin\ \asc)} & \colhead{RA} & \colhead{Dec.} &
\colhead{($^m$)} & \colhead{Name} & \colhead{No.} & \colhead{Name} & \colhead{(AU)} & \colhead{Code}}
\startdata
549.01  & 17:57:48.50 &  +04:41:36.2 &   $-$797.84 & 10326.93 &    9.54 &  Gl 699      &  BD+04 3561 &  G140$-$024 & & HHH \\
419.10  & 10:56:22.46 &  +07:00:27.6 &  $-$3846.7 & $-$2693.5 &   13.46 &  Gl 406      &              &  G045$-$020 & & YYY \\
392.40  & 11:03:20.19 &  +35:58:11.6 &   $-$580.20 & $-$4767.09 &    7.49 &  Gl 411      &  BD+36 2147  &  G119$-$052 & & HHH \\
379.21A & 06:45:08.92 &  $-$16:42:58.0 &   $-$546.01 &  $-$1223.08 &   $-$1.44 &  Gl 244A   &  BD$-$16 1591  &           &  & HHH \\
379.21B & & & & &    8.44 &  Gl 244B   &              &           &   19.7  & HHC \\
373.70A & 01:39:01.74 & $-$17:56:58.5 &   3316.8 &  584.8 &   12.52 &  Gl  65A   &              &  G272$-$061 &  & YYY \\
373.70B & & & & &   12.56 &  Gl  65B   &              &           &    5.1  & YYY \\
336.48  & 18:49:49.36 &  $-$23:50:10.4 &    637.55  & $-$192.47 &   10.37 &  Gl 729      &              &           & & HHH \\
316.00  & 23:41:56.69 &  +44:09:34.5 &   84.6 & $-$1614.8 &   12.27 &  Gl 905      &              &  G171$-$010 & & YYY \\
310.75  & 03:32:55.84 &  $-$09:27:29.7 &   $-$976.44  &   17.97 &    3.72 &  Gl 144      &  BD$-$09  697  &           & & HHH \\
299.58  & 11:47:44.40 &  +00:48:16.4 &    605.62 & $-$1219.23 &   11.12 &  Gl 447      &              &  G010$-$050 & & HHH \\
289.50A & 22:38:37.25 & $-$15:17:07.1 &   2379.8 &  2219.2 &   12.32 &  Gl 866A   &              &  G156$-$031 &  & YYY \\
289.50B & & & & &         &  Gl 866B   &              &  G156$-$031 B &    1.2  & YY  \\
289.50C & & & & &         &  Gl 866C   &              &  G156$-$031 C &    0.3  & YY  \\
287.13A & 21:06:53.94 &  +38:44:57.9 &   4155.10 &  3258.90 &    5.20 &  Gl 820A   &  BD+38 4343  &           &  & HHH \\
285.42B & 21:06:55.26 &  +38:44:31.4 &   4107.40 &  3143.72 &    6.05 &  Gl 820B   &  BD+38 4344  &           &   85.2  & HHH \\
285.93A & 07:39:18.12 &  +05:13:30.0 &   $-$716.57 & $-$1034.58 &    0.40 &  Gl 280A   &  BD+05 1739  &           &  & HHH \\
285.93B & & & & &   10.7  &  Gl 280B   &              &           &   15.9  & HHC \\
280.28A & 18:42:46.69 &  +59:37:49.4 &  $-$1326.88  & 1802.12 &    8.94 &  Gl 725A   &  BD+59 1915  &  G227$-$046 &  & HHH \\
284.48B & 18:42:46.90 &  +59:37:36.6 &  $-$1393.20  & 1845.73 &    9.70 &  Gl 725B   &              &  G227$-$047 &   48.5  & HHH \\
280.27A & 00:18:22.89 &  +44:01:22.6 &   2888.92  &  410.58 &    8.09 &  Gl  15A   &  BD+43   44  &  G171$-$047 &  & HHH \\
280.27B & & & & &   11.06 &  Gl  15B   &              &  G171$-$048 &  155.0  & HHC \\
275.80  & 08:29:44.30 &  +26:46:01.4 &  $-$1139.0 & $-$605.6 &   14.81 &  GJ 1111     &              &  G051$-$015 & & YYY \\
274.17  & 01:44:04.08 &  $-$15:56:14.9 &  $-$1721.82 &   854.07 &    3.49 &  Gl  71      &  BD$-$16  295  &           & & HHH \\
269.05  & 01:12:30.64 &  $-$16:59:56.3 &   1210.09 &   646.95 &   12.10 &  Gl  54.1    &              &  G268$-$135 & & HHH \\
263.26  & 07:27:24.50 &  +05:13:32.8 &    571.27 & $-$3694.25 &    9.84 &  Gl 273      &  BD+05 1668  &  G089$-$019 & & HHH \\
249.52A & 22:27:59.47 &  +57:41:45.1 &   $-$870.23 &  $-$471.10 &    9.59 &  Gl 860A   &  BD+56 2783  &  G232$-$075 &  & HHH \\
249.52B & & & & &    9.85 &  Gl 860B   &              &           &    9.5  & HHY \\
242.89A & 06:29:23.40 &  $-$02:48:50.3 &    694.73 &  $-$618.62 &   11.12 &  Gl 234A   &              &  G106$-$049 &  & HHH \\
242.89B & & & & &   14.6  &  Gl 234B   &              &           &    4.2  & HHC \\
235.24A & 14:49:32.61 &  $-$26:06:20.5 &  $-$1389.70 &   135.76 &   11.72 &  Gl 563.2A   &  CD$-$2510553  &           & & HHH \\
221.80B  & 14:49:31.76 &  $-$26:06:42.0 &  $-$1421.60 &  $-$203.60 &   12.07 &  Gl 563.2B   &              &           & 5.4 & HHH \\
234.51  & 16:30:18.06 &  $-$12:39:45.3 &    $-$93.61 & $-$1184.90 &   10.10 &  Gl 628      &  BD$-$12 4523  &  G153$-$058 & & HHH \\
227.90A & 12:33:16.37 &   +09:01:16.1 &  $-$1797.5 &  220.7 &   12.44 &  Gl 473A   &              &  G012$-$043 &  & YYY \\
227.90B & & & & &   13.04 &  Gl 473B   &              &           &    5.4  & YYY \\
227.45  & 03:22:05.50 &  $-$13:16:43.8 &   $-$112.94 &  $-$299.04 &   12.16 &              &              &  HIP  15689 & & HHH \\
226.95  & 00:49:09.90 &  +05:23:19.0 &   1233.05 & $-$2710.56 &   12.37 &  Gl  35      &              &  G001$-$027 & & HHH \\
224.80  & 02:00:05.90 &  +13:00:34.2 &   1111.2 & $-$1778.4 &   12.26 &  Gl  83.1    &              &  G003$-$033 & & YYY \\
224.00A &  08:58:56.10 &  +08:28:28.0 &   329.1 & $-$320.0 &   10.89 &  1405A   &              &  G041$-$014A &  & GCC \\
224.00B & & & & &         &  1405B   &              &  G041$-$014B &    0.5  & GC  \\
224.00C & & & & &         &  1405C   &              &  G041$-$014C &    3.2  & GC  \\
220.85  & 17:36:25.90 &  +68:20:20.9 &   $-$320.47 & $-$1269.55 &    9.15 &  Gl 687      &  BD+68  946  &  G240$-$063 & & HHH \\
220.30  & 10:48:15.29 & $-$11:21:30.5 &   616.2 & $-$1525.2 &   15.60 &  1679     &              &           & & YYY \\
220.20A & 19:53:56.51 &  +44:24:14.6 &   439.9 & $-$583.8 &   13.41 &  GJ 1245A   &              &  G208$-$044 &  & YYY \\
220.20B & & & & &   14.01 &  GJ 1245B   &              &  G208$-$045 &   47.0  & YYC \\
220.20C & & & & &   13.41 &  GJ 1245C   &              &  G208$-$044 B &    3.7  & YYC \\
213.00  &  00:06:39.52 &  $-$07:34:18.5 &  $-$830.1 & $-$1864.5 &   13.74 &  GJ 1002     &              &  G158$-$027 & & YYY \\
212.69A & 22:53:16.73 &  $-$14:15:49.3 &    960.33 &  $-$675.64 &   10.16 &  Gl 876A   &  BD$-$15 6290  &  G156$-$057 A &  & HHH \\
212.69B & 22:53:16.73 &  $-$14:15:49.3 &    960.33 &  $-$675.64 &         &  Gl 876B   &  BD$-$15 6290  &  G156$-$057 B &    0.21 & HH  \\
206.94A & 11:05:28.58 &  +43:31:36.4 &  $-$4410.79 &   943.32 &    8.82 &  Gl 412A   &  BD+44 2051  &  G176$-$011 &  & HHH \\
206.94B & & & & &   14.40 &  Gl 412B   &              &  G176$-$012 &  190.0  & HHC \\
205.22  & 10:11:22.14 &  +49:27:15.3 &  $-$1361.55 &  $-$505.00 &    6.60 &  Gl 380      &  BD+50 1725  &  G196$-$009 & & HHH \\
204.60  & 10:19:36.23 &  +19:52:10.7 &  $-$503.2 & $-$52.9 &    9.40 &  Gl 388      &  BD+20 2465  &  G054$-$023 & & YYY \\
202.69  & 17:29:36.25 &  +24:39:14.7 &     97.33 &   348.92 &   11.39 &              &              &  HIP  85605 & & HHH \\
198.24A & 04:15:16.32 &  $-$07:39:10.3 &  $-$2239.33 & $-$3419.86 &    4.43 &  Gl 166A   &  BD$-$07  780  &           &  & HHH \\
198.24B & &  &  & &    9.52 &  Gl 166B   &  BD$-$07  781  &  G160$-$060 &  507.5  & HHC \\
198.24C & & & & &   11.17 &  Gl 166C   &              &           &   44.5  & HHC \\
198.00A  & 22:46:48.50 &  +44:19:50.6 &  $-$772.3 & $-$464.0 &   10.06 &  Gl 873A     &  BD+43 4306  &           & & YYY \\
198.07B  & 22:46:49.73 &  +44:20:02.4 &   $-$704.66 &  $-$459.39 & 10.29 &  Gl 873B      &  BD+43 4305  &  G216$-$016 & 167.4 & HHH \\
196.62A & 18:05:27.29 &  +02:30:00.4 &    124.56 &  $-$962.66 &    4.03 &  Gl 702A   &    BD+02 3482  &           &  & HHH \\
196.62B &  &   &     &  &    4.20 &  Gl 702B   &              &           &   22.9  & HHY \\
194.44  & 19:50:47.00 &  +08:52:06.0 &    536.82 &   385.54 &    0.76 &  Gl 768      &  BD+08 4236  &           & & HHH \\
191.86A & 00:15:28.11 &  $-$16:08:01.7 &    728.18 &  $-$617.48 &   11.49 &  GJ 1005A   &              &  G158$-$050 A &  & HHH \\
191.86B &  &  &    &  &         &  GJ 1005B   &              &  G158$-$050 B &    3.9  & HH  \\
191.20A & 08:58:12.21 &  +19:45:45.9 &  $-$873.5 & $-$30.5 &   14.06 &  GJ 1116A   &              &  G009$-$038 &  & YYY \\
191.20B &  &   &  &  &   14.92 &  GJ 1116B   &              &           &   23.6  & YYC \\
186.20  & 06:00:03.23 &  +02:42:15.6 &   229.2 & $-$74.5 &   11.33 &  0999     &              &  G099$-$049 & & YYY \\
185.48  & 11:47:41.38 &  +78:41:28.2 &    743.21 &   480.40 &   10.80 &  Gl 445      &              &  G254$-$029 & & HHH \\
184.13  & 13:45:43.78 &  +14:53:29.5 &   1778.46 & $-$1455.52 &    8.46 &  Gl 526      &  BD+15 2620  &  G063$-$053 & & HHH \\
182.15  & 20:52:33.02 &  $-$16:58:29.1 &   $-$306.70 &    30.78 &   11.41 &              &              &  HIP 103039 & & HHH \\
181.36A & 04:31:11.52 &  +58:58:37.5 &   1300.21 & $-$2048.99 &   10.82 &  Gl 169.1A   &              &  G175$-$034 &  & HHH \\
181.36B &  &   &   &  &   12.44 &  Gl 169.1B   &              &  G175$-$034 &   46.6  & HHC \\
181.32  & 06:54:48.96 &  +33:16:05.4 &   $-$729.33 &  $-$399.31 &    9.89 &  Gl 251      &              &  G087$-$012 & & HHH \\
177.46  & 10:50:52.06 &  +06:48:29.3 &   $-$804.40 &  $-$809.60 &   11.64 &  Gl 402      &              &  G044$-$040 & & HHH \\
175.72  & 05:31:27.40 &  $-$03:40:38.0 &    763.05 & $-$2092.89 &    7.97 &  Gl 205      &  BD$-$03 1123  &  G099$-$015 & & HHH \\
173.41  & 19:32:21.59 &  +69:39:40.2 &    598.43 & $-$1738.81 &    4.67 &  Gl 764      &  BD+69 1053  &           & & HHH \\
173.19A & 06:10:34.62 &  $-$21:51:52.7 &   $-$137.01 &  $-$714.06 &    8.15 &  Gl 229A   &  BD$-$21 1377  &           &  & HHH \\
173.19B & &  &   &   &         &  Gl 229B   &              &             &   48.8  & HH  \\
172.78  & 05:42:09.27 &  +12:29:21.6 &   1999.05 & $-$1570.64 &   11.56 &  Gl 213      &              &  G102$-$022 & & HHH \\
170.26A & 19:16:55.26 &  +05:10:08.1 &   $-$578.86 & $-$1331.70 &    9.12 &  Gl 752A   &  BD+04 4048  &  G022$-$022 &  & HHH \\
170.26B &  &   &   &  &   17.52 &  Gl 752B   &              &           &  521.6  & HHC \\
169.90  & 08:12:41.57 & $-$21:33:11.6 &   37.0 & $-$706.0 &   12.10 &  Gl 300      &              &           & & YYY \\
169.32A & 14:57:28.00 &  $-$21:24:55.7 &   1034.18 & $-$1725.60 &    5.72 &  Gl 570A   &  BD$-$20 4125  &           &  & HHH \\
163.63B & 14:57:26.54 &  $-$21:24:41.5 &    987.05 & $-$1666.81 &    8.01 &  Gl 570B   &  BD$-$20 4123  &           &  107.5  & HHH \\
163.63C & 14:57:26.54 &  $-$21:24:41.5 &    987.05 & $-$1666.81 &         &  Gl 570C   &              &  &    0.8  & HH  \\
169.32D & 14:57:28.34 &  $-$21:26:13.5 &    987.05 & $-$1666.81 &         &  Gl 570D   &              &  &   1400  & HH  \\
168.59  & 07:44:40.17 &  +03:33:08.8 &   $-$344.87 &  $-$450.84 &   11.19 &  Gl 285      &              &  G050$-$004 & & HHH \\
167.99A & 00:49:06.29 &  +57:48:54.7 &   1087.11 &  $-$559.65 &    3.46 &  Gl  34A   &  BD+57  150  &           &  & HHH \\
167.99B & &  &   &   &    7.51 &  Gl  34B   &              &           &   71.00 & HHC \\
167.51  & 23:49:12.53 &  +02:24:04.4 &    995.12 &  $-$968.25 &    8.98 &  Gl 908      &  BD+01 4774  &  G029$-$068 & & HHH \\
167.08A & 17:15:20.98 &  $-$26:36:10.2 &   $-$473.69 & $-$1143.93 &    4.33 &  Gl 663A   &  CD$-$2612026  &           &  & HHH \\
167.08B & 17:15:20.98 &  $-$26:36:10.2 &   $-$473.69 & $-$1143.93 &    5.11 &  Gl 663B   &              &           &   74.0  & HHC \\
167.56C & 17:16:13.36 &  $-$26:32:46.1 &   $-$479.71 & $-$1123.37 &    6.33 &  Gl 664      &  CD$-$2612036  &           & 6390.0  & HHH \\
164.70  & 17:47:51.02 &  +70:53:46.3 &  $-$1246.0  & 1083.2 &   14.15 &  GJ 1221     &              &  G240$-$072 & & YYY \\
163.51  & 14:34:16.81 &  $-$12:31:10.4 &   $-$357.50  &  595.12 &   11.32 &  Gl 555      &  BD$-$11 3759  &           & & HHH \\
163.00  &  05:01:57.60 &  $-$06:56:42.0 &  $-$541.8 & $-$551.3 &   12.1  &  0855     &              &           & & GCC \\
162.50  & 12:13:25.87 &  +10:15:43.5 &  $-$92.8 & $-$14.7 &    5.85 &  1922     &  BD+11 2440  &           & & YYY \\
162.00A & 07:36:25.00 &   +07:04:44.0 &   203.0 & $-$290.0 &   13.22 &  1201A   &              &  G089$-$032 A &  & GCC \\
162.00B & &  &    &  &   13.22 &  1201B   &              &  G089$-$032 B &    3.5  & GCC \\
161.77A & 17:46:14.41 &  $-$32:06:08.3 &    $-$77.62  & $-$270.12 &   11.39 &              &              &  HIP  86963A &  & HHH \\
161.77B & 17:46:12.63 &  $-$32:06:12.8 &    $-$49.82  & $-$319.82 &   10.49 &              &              &  HIP  86963B &  155.0  & HHH \\
161.77C & 17:46:14.41 &  $-$32:06:08.3 &    $-$77.62  & $-$270.12 &         &              &              &  HIP  86963C  &   $<$0.1  & HH  \\
161.59A & 09:14:22.79 &  +52:41:11.8 &  $-$1533.58  & $-$562.80 &    7.64 &  Gl 338A   &  BD+53 1320  &  G195$-$017 &  & HHH \\
159.48B & 09:14:24.70 &  +52:41:11.0 &  $-$1551.30  & $-$656.25 &    7.70 &  Gl 338B   &  BD+53 1321  &  G195$-$018 &  102.0  & HHH \\
160.06A & 23:31:52.18 &  +19:56:14.1 &    554.40  &  $-$62.61 &   10.05 &  Gl 896A   &  BD+19 5116  &  G068$-$024 &  & HHH \\
160.06B &  &  &    &  &   12.4  &  Gl 896B   &              &           &   25.8  & HHC \\
160.06C & &    &    &  &    &  Gl 896C   &              &           &   $<$0.1  & HH \\
160.06D & &   &    &  &         &  Gl 896D   &              &           &   $<$0.1  & HH  \\
159.52  & 15:19:26.82 &  $-$07:43:20.2 &  $-$1224.55  &  $-$99.52 &   10.57 &  Gl 581      &  BD$-$07 4003  &  G151$-$046 & & HHH \\
158.17A & 17:12:07.89 &  +45:39:57.5 &    325.96 & $-$1591.73 &    9.31 &  Gl 661A   &  BD+45 2505  &  G203$-$051 &  & HHH \\
158.17B & &  &   &  &    9.96 &  Gl 661B   &              &           &    4.4  & HHY \\
157.24A & 07:10:01.83 &  +38:31:46.1 &   $-$439.68  & $-$948.36 &   11.65 &  Gl 268A    &              &  G087$-$026 &  & HHH \\
157.24B & &  &   & &         &  Gl 268B    &              &  &    0.1  & HH  \\
156.30  & 14:56:38.58 & $-$30:10:33.6 &  $-$482.5 & $-$835.7 &   17.05 &  2363     &              &           & & YYY \\
156.00  & 13:07:04.30 &  +20:48:38.0 &  $-$71.7 & $-$39.8 &   12.58 &  GJ 2097     &              &           & & GCC \\
155.00  &  05:55:09.37 &  $-$04:10:04.6 &   534.7 & $-$2316.1 &   14.45 &  Gl 223.2    &              &  G099$-$044 & & YYY \\
153.96A & 16:55:28.75 &  $-$08:20:10.8 &   $-$829.34 &  $-$878.81 &    9.02 &  Gl 644A   &  BD$-$08 4352  &           &  & YHH \\
153.96B & 16:55:28.75 &  $-$08:20:10.8 &   $-$829.34 &  $-$878.81 &    9.69 &  Gl 644B   &              &           &    1.4  & YHY \\
153.96C & 16:55:31.75 &  $-$08:23:38.8 &   $-$829.34 &  $-$878.81 &   16.78 &  Gl 644C   &              &           & 1765.0  & YHC \\
153.96D & 16:55:28.75 &  $-$08:20:10.8 &   $-$829.34 &  $-$878.81 &         &  Gl 644D   &              &            &  0.1   & YHH  \\
153.96E & 16:55:25.23 &  $-$08:19:21.3 &   $-$813.47 &  $-$895.23 &   11.73 &  Gl 643      &              &           &  557.8  & HHH \\
153.24  & 23:13:16.98 &  +57:10:06.1 &   2074.37 &   294.97 &    5.57 &  Gl 892      &  BD+56 2966  &           & & HHH \\
152.90  & 12:18:54.77 &  +11:07:41.4 &  $-$1285.0  & 203.5 &   13.79 &  GJ 1156     &              &  G012$-$030 & & YYY \\
151.93  & 16:25:24.62 &  +54:18:14.8 &    432.29 &  $-$170.71 &   10.13 &  Gl 625      &              &  G202$-$048 & & HHH \\
150.96  & 11:00:04.26 &  +22:49:58.7 &   $-$426.31 &  $-$279.94 &   10.03 &  Gl 408      &              &  G058$-$032 & & HHH \\
149.26A & 14:51:23.38 &  +19:06:01.7 &    152.81 &   $-$71.28 &    4.54 &  Gl 566A   &  BD+19 2870  &           &  & HHH \\
149.26B & & &   & &    6.97 &  Gl 566B   &              &           &   32.7  & HHC \\
148.29A & 21:29:36.81 &  +17:38:35.8 &   1008.09 &   376.21 &   10.33 &  Gl 829A     &              &  G126$-$004A &  & HHH \\
148.29B & & &    &    &         &  Gl 829B   &     &  G126$-$004B &    0.5  & HH  \\
146.30  &  08:12:03.21 &   +08:44:41.3 &   1172.2 & $-$5077.4 &   12.83 &  Gl 299      &              &  G050$-$022 & & YYY \\
145.27  & 22:56:34.81 &  +16:33:12.4 &  $-$1033.21  & $-$283.33 &    8.68 &  Gl 880      &  BD+15 4733  &  G067$-$037 & & HHH \\
143.45A & 17:18:57.18 &  $-$34:59:23.3 &   1149.24  &  $-$90.80 &    5.91 &  Gl 667A   &  CD$-$3411626  &           &  & HHH \\
143.45B & & &    &  &    6.27 &  Gl 667B   &              &           &   5.0    & HHY \\
143.45C & &  &     & &   10.24 &  Gl 667C   &              &           &    34.0   & HHC \\
141.95  & 20:53:19.79 &  +62:09:15.8 &      1.08  & $-$774.24 &    8.55 &  Gl 809      &  BD+61 2068  &  G231$-$019 & & HHH \\
138.72A & 02:36:04.89 &  +06:53:12.7 &   1806.27  & 1442.50 &    5.79 &  Gl 105A   &  BD+06  398  &  G073$-$070A &  & HHH \\
138.72B & 02:36:14.20 & +06:52:06.0 &   1806.27  & 1442.50 &    5.82 &  Gl 105B   &              &  G073$-$071 & 1588.0  & HGY \\
138.72C & 02:36:04.89 &  +06:53:12.7 &   1806.27  & 1442.50 &     &  Gl 105C   &  BD+06  398B &  G073$-$070B &   28.8  & HH \\
138.30  & 23:35:14.37 &  $-$02:24:12.6 &   789.1 & $-$846.2 &   14.69 &  GJ 1286     &              &  G157$-$077 & & YYY \\
138.29  & 10:28:55.55 &  +00:50:27.6 &   $-$602.32 &  $-$731.87 &    9.65 &  Gl 393      &  BD+01 2447  &  G055$-$024 & & HHH \\
137.84A & 17:09:31.54 &  +43:40:52.9 &    333.92  & $-$278.02 &   11.77 &  2708A     &              &  G203$-$047A &  & HHH \\
137.84B &  & &   & &         &  2708B   &              &  G203$-$047B &    3.0  & HH  \\
137.50  & 18:19:02.56 &  +66:11:06.1 &   470.2 & $-$408.7 &   13.46 &  2897     &              &  G258$-$033 & & YYY \\
135.30A & 00:24:41.30 & $-$27:08:52.8 &   $-$53.5 &  611.7 &   15.42 &  GJ 2005A   &              &           &  & YYY \\
135.30B &  & & & &    &  GJ 2005B   &              &           &    7.39 & YY \\
135.30C &  & &   &  &    &  GJ 2005C   &              &           &   14.80 & YY \\
134.40  & 22:23:07.54 & $-$17:37:01.1 &   291.4 & $-$721.3 &   13.25 &  3517     &              &           & & YYY \\
134.04  & 00:48:22.98 &  +05:16:50.2 &    758.04 & $-$1141.22 &    5.74 &  Gl  33      &  BD+04  123  &           & & HHH \\
133.91  & 01:42:29.76 &  +20:16:06.6 &   $-$302.12 &  $-$677.40 &    5.24 &  Gl  68      &  BD+19  279  &           & & HHH \\
132.60  & 18:07:30.80 & $-$15:58:14.2 &  $-$563.1 &  $-$351.9 &   13.64 &  GJ 1224     &              &  G154$-$044 & & YYY \\
132.42  & 02:44:15.51 &  +25:31:24.1 &    864.77 &  $-$367.17 &   10.55 &  Gl 109      &              &  G036$-$031 & & HHH \\
132.40A & 01:08:16.39 &  +54:55:13.2 &   3421.44 & $-$1599.27 &    5.17 &  Gl  53A   &  BD+54  223  &           &  & HHH \\
132.40B & &  &    & &   11.   &  Gl  53B   &              &           &    4.8  & HHC \\
132.10  & 06:01:11.30 &  +59:35:00.2 &  $-$144.3 & $-$818.4 &   11.71 &  998     &              &  G192$-$013 & & YYY \\
131.12  & 13:29:59.79 &  +10:22:37.8 &   1128.00 & $-$1074.30 &    9.05 &  Gl 514      &  BD+11 2576  &  G063$-$034 & & HHH \\
130.94  & 22:56:24.05 &  $-$31:33:56.0 &    330.53 &  $-$159.86 &    6.48 &  Gl 879      &  CD$-$3217321  &           & & HHH \\
130.08  & 22:57:39.05 &  $-$29:37:20.1 &    329.22 &  $-$164.22 &    1.17 &  Gl 881      &  CD$-$3019370  &           & & HHH \\
129.54  & 17:25:45.23 &  +02:06:41.1 &   $-$580.47 & $-$1184.81 &    7.54 &  Gl 673      &  BD+02 3312  &  G019$-$024 & & HHH \\
129.40A & 05:02:28.42 & $-$21:15:23.9 & $-$141.55 & $-$221.74 &    8.31 &  Gl 185  A   &  BD-21 1051  &           &  & YHH \\
129.40B & & &  &  &         &  Gl 185  B   &         &   &    9.5  & YH  \\
128.93  & 18:36:56.34 &  +38:47:01.3 &    201.02 &   287.46 &    0.03 &  Gl 721      &  BD+38 3238  &           & & HHH \\
128.80  & 06:59:32.00 &  +19:19:55.4 &   835.4  & $-$895.9 &   14.83 &  GJ 1093     &              &  G109$-$035 & & YYY \\
128.28  & 18:05:07.58 &  $-$03:01:52.7 &    570.14 &  $-$332.59 &    9.37 &  Gl 701      &  BD$-$03 4233  &  G020$-$022 & & HHH \\
127.99  & 10:12:17.67 &  $-$03:44:44.4 &   $-$152.93 &  $-$242.90 &    9.26 &  Gl 382      &  BD$-$03 2870  &  G053$-$029 & & HHH \\
126.00  & 13:31:46.70 &  +29:16:36.0 &  $-$227.7 & $-$159.5 &   11.95 &  2128     &              &  G165$-$008 & & GCC \\
125.62  & 20:30:32.05 &  +65:26:58.4 &    443.25  &  284.06 &   10.54 &  Gl 793      &              &  G262$-$015 & & HHH \\
125.00  & 05:56:23.69 &  +05:20:46.9 &  $-$479.4 & $-$940.9 &   14.11 &  GJ 1087     &              &  G099$-$047 & & YYY \\
\enddata
\tablecomments{The coordinates are given in the J2000.00 equinox and epoch,
meaning that they include the proper motion and precession.  Source
Codes: G: \citet{JGthesis}, C: CNS3, H: Hipparcos, Y: Yale.  The three
characters correspond to parallax, position and V band magnitude
respectively and are meant to indicate where the listed value comes
from.\\ SMA means semi$-$major axis and is derived from the projected
separation if the companion is visibly detected or the radial velocity
orbital solution (\cite{reid97}).  Positions for companions are only
given if grossly different from the primary star position or if needed
for identification purposes (i.e., to establish which star the
companion is closest to in a multiple star system).}

\end{deluxetable}

\clearpage

\begin{deluxetable}{lrll}
\tablecolumns{4}
\tabletypesize{\footnotesize}
\tablewidth{405pt}
\tablecaption{List of Observations of the 8 pc Sample\label{chap4:tab2}}
\tablehead{
\colhead{Para.} & \colhead{V\phantom{i}} & \colhead{Dates of}  & \colhead{Dates of} \\
\colhead{(mas)} & \colhead{($^m$)} & \colhead{AOC Observations} & \colhead{Infrared Observations}}
\startdata
549.01 &    9.54 & 9/92; 4/94 & 8/96; 7/97 \\
419.10 &   13.46 & 2/97; 4/97 & 12/96 \\
392.40 &    7.49 & 4/94 & 12/96 \\
379.21 &   -1.44 & 11/96; 1/98 & 10/96  \\
373.70 &   12.52 & 10/94; 10/95b & 8/96 \\
336.48 &   10.37 & 6/95; 8/95; 10/95a; 6/96 & 8/96; 7/97 \\
316.00 &   12.27 & 9/92; 8/95; 10/95a; 10/95b; 9/97~~~~~~~~~~ & 8/96; 7/97 \\
310.75 &    3.72 & 10/95b & 8/96; 11/97 \\
299.58 &   11.12 & 1/93; 2/96; 4/97 & 12/96 \\
289.50 &   12.32 & 6/95; 8/95; 9/97 & 9/95; 8/96 \\
287.13 &    5.20 & 10/95b; 8/96 & 8/96 \\
285.93 &    0.40 & 2/95; 11/96; 1/98 & 11/97  \\
280.28 &    8.94 & 6/92; 8/96; 4/97 & 8/96; 7/97 \\
280.27 &    8.09 & 10/95b; 9/97 & 8/96; 12/96; 8/97; \\
275.80 &   14.81 & too faint & 11/95; 12/96 \\
274.17 &    3.49 & 10/95b; 11/96 & 11/95; 8/96; 8/97 \\
269.05 &   12.10 & 8/95; 11/96; 9/97 & 9/95; 8/96; \\
263.26 &    9.84 & 2/95; 2/97 & 12/96; 11/97 \\
249.52 &    9.59 & 10/95b; 8/96 & 11/95; 7/97 \\
242.89 &   11.12 & 2/95; 12/95; 11/96 & 11/95; 10/96; 12/96 \\
235.24 &   11.72 &  & 3/99\\
234.51 &   10.10 & 4/94; 4/97; 6/97 & 8/96; 7/97 \\
227.90 &   12.44 & 1/93; 2/96; 4/97 & 12/96 \\
227.45 &   12.16 &  & 3/99\\
226.95 &   12.37 & 9/92; 8/95 & 8/96; 12/96 \\
224.80 &   12.26 & 8/95; 10/95a; 10/95b; 9/97 & 9/95; 11/95; 8/96; 10/96~~~~~~~~~~~~~~ \\
224.00 &   10.89 & 2/95; 11/95; 11/96; 1/98; 3/98 & 12/96; 3/98; 12/98 \\
221.80 &   12.07 &  & 3/99\\
220.85 &    9.15 & 6/95; 8/95; 9/97 & 8/96; 7/97 \\
220.30 &   15.60 & too faint & 2/96; 12/96 \\
220.20 &   13.41 & 9/92; 8/97 & 11/95; 8/96; 7/97; 8/97 \\
213.00 &   13.74 & 9/97 & 10/96 \\
212.69 &   10.16 & 10/93; 10/94; 8/95; 9/97 & 8/96; 7/97 \\
206.94 &    8.82 & 4/94 & 2/96; 12/96 \\
205.22 &    6.60 & 1/93; 4/97 & 2/96; \\
204.60 &    9.40 & 4/94; 2/95; 12/95 & 12/95; 12/96 \\
202.69 &   11.39 &  & 3/99\\
198.24 &    4.43 & 2/96; 11/96; 9/97 & 10/96; 12/96; \\
198.00 &   10.06 &  & \\
196.62 &    4.03 & 8/96; 8/97; 9/97 & 8/96  \\
194.44 &    0.76 & 10/95b & 8/96 \\
191.86 &   11.49 & 11/96; 9/97 & 12/96; 8/97 \\
191.20 &   14.06 & too faint & 12/96; 3/98 \\
186.20 &   11.33 & 12/95; 11/96; 1/98 & 10/96; 12/96; 11/97 \\
185.48 &   10.80 & 2/97 & 12/96 \\
184.13 &    8.46 & 1/93; 4/94; 4/97 & 12/96; 7/97 \\
182.15 &   11.41 &  & \\
181.36 &   10.82 & 11/96; 9/97 & 10/96; 12/96 \\
181.32 &    9.89 & 11/96; 1/98 & 10/96 \\
177.46 &   11.64 & 2/97; 4/97; 1/98 & 12/96 \\
175.72 &    7.97 & 11/96; 1/98 & 12/96  \\
174.23 &    9.02 & 8/96; 4/97 & 8/96; 7/97 \\
173.41 &    4.67 & 8/96 & 8/96 \\
173.19 &    8.15 & 10/94; 2/95; 10/95a & 9/95; 11/95; 10/96; 12/96~~~~~~~~~~~~~~~~~ \\
172.78 &   11.56 & 11/96; 1/98 & 10/96; 12/96; 11/97 \\
170.26 &    9.12 & 9/92; 6/96 & 8/96; 7/97 \\
169.90 &   12.10 & 11/96; 1/98 & 12/96; 11/97  \\
169.32 &    5.72 & 4/97 & 7/97 \\
168.59 &   11.19 & 2/95; 11/95; 11/96; 1/98~~~~~~~~~~~~~~~~~ & 12/95;  12/96; 11/97 \\
167.99 &    3.46 & 11/96; 8/97 & 10/96; 8/97 \\
167.51 &    8.98 & 9/92; 10/93; 11/96; 9/97 & 8/96; 8/97 \\
167.08 &    4.33 & 4/97; 9/97 & 7/97; 6/98 \\
164.70 &   14.15 & too faint & 3/98 \\
163.51 &   11.32 & 6/93; 4/97 & 7/97 \\
163.00 &   12.1  & 11/96 & 10/96; 12/96 \\
162.50 &    5.85 & 4/97 & 12/96 \\
162.00 &   13.22 & 1/98 & 12/95; 12/96 \\
161.77 &   11.39 &  & 3/99\\
161.59 &    7.64 & 11/96 & 12/96 \\
160.06 &   10.05 & 10/95a; 9/97 & 11/95; 8/96; 7/97 \\
159.52 &   10.57 & 6/93; 4/94 & 7/97; 3/98 \\
158.17 &    9.31 & 4/94; 9/97 &  8/96; 7/97 \\
157.24 &   11.65 & 10/94; 11/95 & 11/95; 10/96 \\
156.30 &   17.05 & too faint & 3/98; 6/98 \\
156.00 &   12.58 & 4/97; 6/97 & 12/96\\
155.00 &   14.45 & too faint & 10/96; 12/96; 3/98 \\
153.24 &    5.57 & 8/96; 11/96; 8/97 & 8/96; 7/97 \\
152.90 &   13.79 & 4/97; 1/98 & 2/96; 12/96 \\
151.93 &   10.13 & 4/97 & 7/97 \\
150.96 &   10.03 & 2/97; 4/97; 1/98 & 12/96\\
149.26 &    4.54 & 4/97 & 7/97 \\
148.29 &   10.33 & 10/94; 8/96 & 8/96 \\
146.30 &   12.83 & 11/96; 4/97 & 12/96; 11/97; 3/98 \\
145.27 &    8.68 & 11/96; 9/97; 1/98 & 10/96; 7/97 \\
143.45 &    5.91 & 8/97; 9/97 & 7/97; 6/98 \\
141.95 &    8.55 & 6/93; 6/96; 11/96 & 8/96; 7/97 \\
138.72 &    5.79 & 10/93; 10/94; 8/95; 10/95a & 9/95; 10/96; 12/96; 8/97; 12/98~~~~~~~~ \\
138.30 &   14.69 & too faint & 10/96; 10/97 \\
138.29 &    9.65 & 4/94; 2/95; 11/96; 4/97; 1/98; 3/98~~~~ & 12/96 \\
137.84 &   11.77 & & 7/97 \\
137.50 &   13.46 & too faint & 7/97 \\
135.30 &   15.42 & too faint & 9/95; 8/96 \\
134.40 &   13.25 & 9/97 & 7/97  \\
134.04 &    5.74 & 10/93; 11/96 & 8/96; 12/96; 8/97 \\
133.91 &    5.24 & 10/93; 8/96; 9/97 & 8/96 \\
132.60 &   13.64 & too faint & 7/97; 6/98 \\
132.42 &   10.55 & 10/95b; 2/97 & 8/96 \\
132.40 &    5.17 & 8/96; 11/96; 8/97 & 12/96  \\
132.10 &   11.71 & 11/96; 1/98 & 10/96; 12/96; 11/97 \\
131.12 &    9.05 & 4/97 & 12/96  \\
130.94 &    6.48 & 8/97 & 10/96; 7/97 \\
130.08 &    1.17 & 11/96; 9/97 & 10/96  \\
129.54 &    7.54 & 4/94; 4/97 & 8/96; 7/97 \\
129.40 &    8.31 & 11/96; 1/98 & 10/96; 12/96; 11/97; 3/98; 3/99 \\
128.93 &    0.03 & 9/97 &  7/97 \\
128.80 &   14.83 & too faint & 10/96; 12/96; 11/97 \\
128.28 &    9.37 & 8/97; 9/97 & 7/97; 3/98; 6/98 \\
127.99 &    9.26 &  & \\
126.00 &   11.95 & 6/95; 4/97; 6/97 & 2/96; 12/96; 7/97 \\
125.62 &   10.54 &  & 3/99\\
125.00 &   14.11 & too faint & 10/96; 11/97 \\
\enddata
\end{deluxetable}

\clearpage

\begin{deluxetable}{ccc}
\tablecolumns{3}
\tabletypesize{\footnotesize}
\tablewidth{0pt}
\tablecaption{Trapezium Calibration Field\label{chap4:tab4}}
\tablehead{\colhead{Star} & \multicolumn{2}{c}{Position (J2000.00)} \\
\colhead{Name} & \colhead{RA ($^h$ $^m$ $^s$)} & \colhead{Dec ($^\circ$ \arcmin\ \asc)}}
\startdata
HR 1895 & 05:35:16.462 & $-$05:23:23.03 \\
HR 1893 & 05:35:15.821 & $-$05:23:14.45 \\
HR 1894 & 05:35:16.129 & $-$05:23:06.96 \\
HR 1896 & 05:35:17.248 & $-$05:23:16.69 \\
FS 1 & 05:35:15.768 & $-$05:23:10.06 \\
FS 2 & 05:35:15.953 & $-$05:23:49.99 \\
\enddata
\tablecomments{HR 1895 is the central, occulted
star in the image.  Position data are from McCaughrean and Stauffer
(1994).}
\end{deluxetable}

\clearpage

\begin{deluxetable}{ccc}
\tablecolumns{3}
\tabletypesize{\footnotesize}
\tablewidth{0pt}
\tablecaption{M5 Calibration Field\label{chap4:tab5}}
\tablehead{\colhead{Star} & \multicolumn{2}{c}{Position (B1950.00)} \\
\colhead{Name} & \colhead{RA ($^h$ $^m$ $^s$)} & \colhead{Dec ($^\circ$ \arcmin\ \asc)}}
\startdata
93 & 15:16:01.70 & +02:11:31.7 \\
94 & 15:16:03.14 & +02:12:00.0 \\
95 & 15:16:02.46 & +02:12:09.6 \\
96 & 15:16:04.36 & +02:12:09.7 \\
97 & 15:16:05.05 & +02:12:20.3 \\
\enddata
\tablecomments{Positions and numbering scheme are from
Cudworth (1979).  Star 93 is occulted and centered in the
image.}
\end{deluxetable}

\clearpage

\begin{deluxetable}{ccc}
\tablecolumns{3}
\tabletypesize{\footnotesize}
\tablewidth{0pt}
\tablecaption{M15 Calibration Field\label{chap4:tab6}}
\tablehead{\colhead{Star} & \multicolumn{2}{c}{Position (B1950.00)} \\
\colhead{Name} & \colhead{RA ($^h$ $^m$ $^s$)} & \colhead{Dec ($^\circ$ \arcmin\ \asc)}}
\startdata
275 & 21:27:22.82 & +11:55:33.6 \\
277 & 21:27:24.40 & +11:55:15.3 \\
278 & 21:27:24.83 & +11:55:30.2 \\
279 & 21:27:25.93 & +11:55:30.0 \\
280 & 21:27:26.80 & +11:55:43.5 \\
281 & 21:27:25.24 & +11:55:51.0 \\
282 & 21:27:24.09 & +11:55:52.1 \\
\enddata
\tablecomments{Positions and numbering scheme are
from Cudworth (1976).  Star 278 is occulted and centered in the
image.}
\end{deluxetable}

\clearpage

\begin{deluxetable}{crrrrrrrrrr}
\tablecolumns{11}
\tabletypesize{\footnotesize}
\tablewidth{0pt}
\tablecaption{Percentage of Survey Stars with Observations Sensitive to Companions of All J Magnitudes\label{tab:Jmag}}
\tablehead{
\colhead{$M_J$} & \multicolumn{10}{c}{Separation (A.U.)} \\
\colhead{} &  
\colhead{2.5} &   
\colhead{5} &   
\colhead{10} &   
\colhead{20} &   
\colhead{40} &   
\colhead{80} &  
\colhead{120} &  
\colhead{160} &  
\colhead{200} &  
\colhead{225}}
\startdata
 9.9 &    1 &   31 &   97 &   99 &   98 &   60 &    2 &    0 &    0 &    0\\
10.9 &    1 &   31 &   97 &   99 &   98 &   60 &    2 &    0 &    0 &    0\\\tableline
11.9 &    1 &   31 &   97 &   99 &   98 &   60 &    2 &    0 &    0 &    0\\
12.9 &    1 &   31 &   97 &   95 &   94 &   59 &    2 &    0 &    0 &    0\\
13.9 &    1 &   26 &   81 &   90 &   83 &   59 &    2 &    0 &    0 &    0\\
14.9 &    0 &    5 &   46 &   80 &   80 &   56 &    2 &    0 &    0 &    0\\
15.9 &    0 &    4 &   28 &   66 &   73 &   48 &    2 &    0 &    0 &    0\\
16.9 &    0 &    1 &    5 &   33 &   65 &   45 &    2 &    0 &    0 &    0\\
17.9 &    0 &    0 &    0 &    4 &   25 &   15 &    1 &    0 &    0 &    0\\
18.9 &    0 &    0 &    0 &    0 &    0 &    0 &    0 &    0 &    0 &    0\\
\enddata
\tablecomments{$M_J$ is the absolute magnitude in J band.  The line above 
the 11.9$^m$ entry is meant to demarcate all stars (above the line) from 
cool brown dwarfs.}
\end{deluxetable}

\clearpage

\begin{deluxetable}{crrrrrrrrrr}
\tablecolumns{11}
\tabletypesize{\footnotesize}
\tablewidth{0pt}
\tablecaption{Percentage of Survey Stars with Observations Sensitive to Companions of All $z$ Magnitudes\label{tab:zmag}}
\tablehead{
\colhead{$M_z$} & \multicolumn{10}{c}{Separation (A.U.)} \\
\colhead{} &  
\colhead{2.5} &   
\colhead{5} &   
\colhead{10} &   
\colhead{20} &   
\colhead{40} &   
\colhead{80} &  
\colhead{120} &  
\colhead{160} &  
\colhead{200} &  
\colhead{225}}
\startdata
11.5 &    0 &    1 &   29 &  100 &  100 &   97 &   84 &   60 &   26 &   13\\
12.5 &    0 &    1 &   29 &  100 &  100 &   97 &   84 &   60 &   26 &   13\\
13.5 &    0 &    1 &   26 &   83 &   99 &   97 &   84 &   60 &   26 &   13\\
14.5 &    0 &    1 &   25 &   75 &   99 &   97 &   84 &   60 &   26 &   13\\\tableline
15.5 &    0 &    1 &   25 &   67 &   98 &   97 &   84 &   60 &   26 &   13\\
16.5 &    0 &    0 &   20 &   59 &   92 &   95 &   84 &   60 &   26 &   13\\
17.5 &    0 &    0 &    3 &   28 &   83 &   95 &   83 &   60 &   26 &   13\\
18.5 &    0 &    0 &    1 &    8 &   67 &   94 &   83 &   59 &   26 &   13\\
19.5 &    0 &    0 &    0 &    0 &    9 &   67 &   66 &   52 &   25 &   12\\
20.5 &    0 &    0 &    0 &    0 &    0 &    0 &    0 &    0 &    0 &    0\\
\enddata
\tablecomments{$M_z$ is the absolute
 magnitude in $z$ band.  The line below the 14.5$^m$ entry demarcates
 all stars (above the line) from cool brown dwarfs.}
\end{deluxetable}

\clearpage

\begin{deluxetable}{crrrrrrrrrr}
\tablecolumns{11}
\tabletypesize{\footnotesize}
\tablewidth{0pt}
\tablecaption{Percentage of Survey Stars with Observations Sensitive to Companions of All $r$ Magnitudes\label{tab:rmag}}
\tablehead{
\colhead{$M_r$} & \multicolumn{10}{c}{Separation (A.U.)} \\
\colhead{} &  
\colhead{2.5} &   
\colhead{5} &   
\colhead{10} &   
\colhead{20} &   
\colhead{40} &   
\colhead{80} &  
\colhead{120} &  
\colhead{160} &  
\colhead{200} &  
\colhead{225}}
\startdata
13.7 &    0 &    1 &   29 &  100 &  100 &   97 &   84 &   60 &   26 &   13\\
14.7 &    0 &    1 &   29 &  100 &  100 &   97 &   84 &   60 &   26 &   13\\
15.7 &    0 &    1 &   29 &   98 &   90 &   95 &   84 &   60 &   26 &   13\\
16.7 &    0 &    1 &   27 &   89 &   85 &   95 &   83 &   60 &   26 &   13\\\tableline
17.7 &    0 &    0 &   20 &   72 &   82 &   93 &   82 &   59 &   25 &   13\\
18.7 &    0 &    0 &    6 &   54 &   79 &   87 &   82 &   59 &   25 &   11\\
19.7 &    0 &    0 &    5 &   51 &   79 &   80 &   77 &   58 &   25 &   11\\
20.7 &    0 &    0 &    0 &    5 &   57 &   77 &   70 &   52 &   25 &   11\\
21.7 &    0 &    0 &    0 &    0 &    8 &   58 &   64 &   44 &   17 &    9\\
22.7 &    0 &    0 &    0 &    0 &    0 &    0 &    0 &    0 &    0 &    0\\
\enddata
\tablecomments{$M_r$ is the absolute
 magnitude in $r$ band.  The line below the 16.7$^m$ entry demarcates
 all stars (above the line) from brown dwarfs.}
\end{deluxetable}

\clearpage

\begin{deluxetable}{crrrrrrrrrr}
\tablecolumns{11}
\tabletypesize{\footnotesize}
\tablewidth{0pt}
\tablecaption{Percentage of Survey Stars with Observations in J, $z$ and $r$ Band Sensitive to 0.08 M$_\odot$ Stellar Companions\label{tab:hbml}}
\tablehead{
\colhead{Band} & \multicolumn{10}{c}{Separation (A.U.)} \\
\colhead{} &  
\colhead{2.5} &   
\colhead{5} &   
\colhead{10} &   
\colhead{20} &   
\colhead{40} &   
\colhead{80} &  
\colhead{120} &  
\colhead{160} &  
\colhead{200} &  
\colhead{225}}
\startdata
 J  &    1 &   31 &   97 &   99 &   98 &   60 &    2 &    0 &    0 &    0\\
$z$ &    0 &    1 &   25 &   72 &   98 &   97 &   84 &   60 &   26 &   13\\
$r$ &    0 &    1 &   27 &   89 &   82 &   95 &   83 &   59 &   26 &   13\\
\enddata
\end{deluxetable}

\clearpage

\begin{deluxetable}{crrrrrrrrrr}
\tablecolumns{11}
\tabletypesize{\footnotesize}
\tablewidth{0pt}
\tablecaption{Percentage of Survey Stars with Observations in J, $z$ and $r$ Band Sensitive to Gliese 229B-like Companions\label{tab:229B}}
\tablehead{
\colhead{Band} & \multicolumn{10}{c}{Separation (A.U.)} \\
\colhead{} &  
\colhead{2.5} &   
\colhead{5} &   
\colhead{10} &   
\colhead{20} &   
\colhead{40} &   
\colhead{80} &  
\colhead{120} &  
\colhead{160} &  
\colhead{200} &  
\colhead{225}}
\startdata
J   &    0 &    5 &   37 &   70 &   77 &   50 &    2 &    0 &    0 &    0\\
$z$ &    0 &    0 &    0 &    5 &   55 &   92 &   82 &   59 &   26 &   13\\
$r$ &    0 &    0 &    0 &    0 &    0 &    0 &    0 &    0 &    0 &    0\\
\enddata
\end{deluxetable}

\clearpage

\begin{deluxetable}{crrrrrrrrrr}
\tablecolumns{11}
\tabletypesize{\footnotesize}
\tablewidth{0pt}
\tablecaption{Percentage of Survey Stars with $z$ band Observations Sensitive to Companions of Brown Dwarf Mass: Age = 5 Gyr\label{tab:zmag5}}
\tablehead{
\colhead{Mass} & \multicolumn{10}{c}{Separation (A.U.)} \\
\colhead{M$_J$} &  
\colhead{2.5} &   
\colhead{5} &   
\colhead{10} &   
\colhead{20} &   
\colhead{40} &   
\colhead{80} &  
\colhead{120} &  
\colhead{160} &  
\colhead{200} &  
\colhead{225}}
\startdata
70 &    0 &    1 &   25 &   67 &   98 &   95 &   84 &   60 &   26 &   13\\
65 &    0 &    0 &   20 &   59 &   93 &   95 &   84 &   60 &   26 &   13\\
59 &    0 &    0 &    4 &   34 &   92 &   95 &   84 &   60 &   26 &   13\\
51 &    0 &    0 &    3 &   28 &   83 &   95 &   83 &   60 &   26 &   13\\
45 &    0 &    0 &    3 &   28 &   82 &   95 &   83 &   60 &   26 &   13\\
39 &    0 &    0 &    1 &    8 &   67 &   93 &   83 &   59 &   26 &   13\\
35 &    0 &    0 &    0 &    1 &   28 &   86 &   82 &   59 &   26 &   13\\
34 &    0 &    0 &    0 &    1 &   24 &   79 &   76 &   59 &   25 &   13\\
33 &    0 &    0 &    0 &    0 &    0 &    0 &    0 &    0 &    0 &    0\\
\enddata
\tablecomments{Here, M$_J$ is the mass of Jupiter.}
\end{deluxetable}

\clearpage

\begin{deluxetable}{crrrrrrrrrr}
\tablecolumns{11}
\tabletypesize{\footnotesize}
\tablewidth{0pt}
\tablecaption{Percentage of Survey Stars with J band Observations Sensitive to Companions of Brown Dwarf Mass: Age = 5 Gyr\label{tab:Jmag5}}
\tablehead{
\colhead{Mass} & \multicolumn{10}{c}{Separation (A.U.)} \\
\colhead{M$_J$} &  
\colhead{2.5} &   
\colhead{5} &   
\colhead{10} &   
\colhead{20} &   
\colhead{40} &   
\colhead{80} &  
\colhead{120} &  
\colhead{160} &  
\colhead{200} &  
\colhead{225}}
\startdata
70 &    0 &    5 &   39 &   77 &   77 &   52 &    2 &    0 &    0 &    0\\
60 &    0 &    4 &   28 &   65 &   73 &   48 &    2 &    0 &    0 &    0\\
50 &    0 &    1 &    9 &   46 &   68 &   46 &    2 &    0 &    0 &    0\\
45 &    0 &    1 &    5 &   33 &   65 &   44 &    2 &    0 &    0 &    0\\
40 &    0 &    1 &    1 &   16 &   50 &   42 &    2 &    0 &    0 &    0\\
35 &    0 &    0 &    0 &    4 &   25 &   17 &    1 &    0 &    0 &    0\\
34 &    0 &    0 &    0 &    0 &    0 &    0 &    0 &    0 &    0 &  0\\
\enddata
\tablecomments{Here,
M$_J$ is the mass of Jupiter.}
\end{deluxetable}

\clearpage

\begin{deluxetable}{crrrrrrrrrr}
\tablecolumns{11}
\tabletypesize{\footnotesize}
\tablewidth{0pt}
\tablecaption{Percentage of Survey Stars with $z$ band Observations Sensitive to Companions of Brown Dwarf Mass: Age = 1 Gyr\label{tab:zmag1}}
\tablehead{
\colhead{Mass} & \multicolumn{10}{c}{Separation (A.U.)} \\
\colhead{M$_J$} &  
\colhead{2.5} &   
\colhead{5} &   
\colhead{10} &   
\colhead{20} &   
\colhead{40} &   
\colhead{80} &  
\colhead{120} &  
\colhead{160} &  
\colhead{200} &  
\colhead{225}}
\startdata
65 &    0 &    1 &   26 &   83 &   99 &   97 &   84 &   60 &   26 &   13\\
60 &    0 &    1 &   26 &   83 &   99 &   97 &   84 &   60 &   26 &   13\\
51 &    0 &    1 &   25 &   74 &   98 &   97 &   84 &   60 &   26 &   13\\
40 &    0 &    1 &   25 &   67 &   98 &   97 &   84 &   60 &   26 &   13\\
31 &    0 &    0 &    4 &   34 &   92 &   95 &   84 &   60 &   26 &   13\\
21 &    0 &    0 &    1 &    8 &   69 &   95 &   83 &   60 &   26 &   13\\
15 &    0 &    0 &    0 &    1 &   28 &   83 &   81 &   59 &   26 &   13\\
14 &    0 &    0 &    0 &    0 &    0 &    0 &    0 &    0 &    0 &    0\\
\enddata
\tablecomments{Here, M$_J$ is the mass of Jupiter.}
\end{deluxetable}

\clearpage

\begin{deluxetable}{crrrrrrrrrr}
\tablecolumns{11}
\tabletypesize{\footnotesize}
\tablewidth{0pt}
\tablecaption{Percentage of Survey Stars with J band Observations Sensitive to Companions of Brown Dwarf Mass: Age = 1 Gyr\label{tab:Jmag1}}
\tablehead{
\colhead{Mass} & \multicolumn{10}{c}{Separation (A.U.)} \\
\colhead{M$_J$} &  
\colhead{2.5} &   
\colhead{5} &   
\colhead{10} &   
\colhead{20} &   
\colhead{40} &   
\colhead{80} &  
\colhead{120} &  
\colhead{160} &  
\colhead{200} &  
\colhead{225}}
\startdata
70 &    1 &   30 &   96 &   94 &   94 &   59 &    2 &    0 &    0 &    0\\
60 &    1 &   30 &   91 &   90 &   83 &   59 &    2 &    0 &    0 &    0\\
50 &    1 &   18 &   64 &   83 &   81 &   58 &    2 &    0 &    0 &    0\\
40 &    0 &    5 &   46 &   80 &   78 &   56 &    2 &    0 &    0 &    0\\
30 &    0 &    4 &   28 &   66 &   74 &   49 &    2 &    0 &    0 &    0\\
25 &    0 &    1 &    9 &   46 &   71 &   46 &    2 &    0 &    0 &    0\\
20 &    0 &    1 &    5 &   31 &   64 &   44 &    2 &    0 &    0 &    0\\
16 &    0 &    1 &    1 &   16 &   48 &   42 &    2 &    0 &    0 &    0\\
13 &    0 &    0 &    0 &    0 &    0 &    0 &    0 &    0 &    0 &    0\\
\enddata
\tablecomments{Here, M$_J$ is the mass of Jupiter.}
\end{deluxetable}


\begin{thebibliography}

\bibitem[Alcock et al.(1998)]{al98} Alcock, C. et al.\
1998, \apjl, 499, L9

\bibitem[Burgasser et al.(2000)]{bur00} Burgasser, A. J. et al. 
2000, \apjl, 531, L57

\bibitem[Burgasser et al.(1999)]{bur99} Burgasser, A. J. et al. 1999, \apj, 
522, 65 

\bibitem[Burrows and Liebert(1993)]{BL93} Burrows, A. and Liebert,
J. 1993, Rev.\ Mod.\ Phys., 65, 301

\bibitem[Burrows et al.(1997)]{B97} Burrows, A., Marley M.,
Hubbard W. B., Lunine J. I., Guillot T., Saumon D., Freedman R.,
Sudarsky D. and Sharp C. 1997, \apj, 491, 856

\bibitem[Burrows et al.(2000)]{b00} Burrows, A., Marley, M. S., and Sharp, C. M. 2000, \apj, 531, 438

\bibitem[Chabrier and M\'{e}ra(1998)]{cm98} Chabrier, G. and
M\'{e}ra, D. 1998, ASP Conf.\ Ser.,134, 495

\bibitem[Cudworth(1976)]{cud76} Cudworth, K. M. 1976, \aj, 81, 519

\bibitem[Cudworth(1979)]{cud79} Cudworth, K. M. 1979, \aj, 84, 1866

\bibitem[Delfosse et al.(1998)]{delfoss98a}Delfosse, X.,
Forveille, T., Mayor, M., Perrier, C., Naef, D. and Queloz, D. 1998,
\aap, 338, L67

\bibitem[Delfosse et al.(1999)]{delfoss98b}Delfosse, X.,
Forveille, T., Beuzit, J.-L., Udry, S., Mayor, M. and Perrier,
C. 1999, \aap, 344, 897

\bibitem[Gizis et al.(2000)]{giz00} Gizis, J. E., Monet, D. G., Reid,
I. N., Kirkpatrick, J. D., Liebert, J. and Williams, R. J. 2000, \aj,
120, 1085

\bibitem[Gliese and Jahreiss(1991)]{GJ91} Gliese, W. and Jahreiss,
             H. 1991, Preliminary Version of the Third Catalog of
             Nearby Stars.

\bibitem[Golimowski et al.(1992)]{G92} Golimowski, D. A., Clampin,
            Durrance, S. T. and Barkhouser, R. H. 1992, Appl.\ Opt.,
            31, 4405

\bibitem[Golimowski et al.(1995)]{gol95} Golimowski, D. A., 
Nakajima, T., Kulkarni, S. R. and Oppenheimer, B. R. 1995, \apjl,
444, L101

\bibitem[Golimowski et al.(1998)]{gol98} Golimowski, D. A., 
Kulkarni, S. R., Burrows, C.  J., Brukardt, R. A. and Oppenheimer,
B. R.  1998, \aj, 115, 2579

\bibitem[Henry and McCarthy(1990)]{hm90}Henry, T. J. and
McCarthy, D. W. Jr.  1990, \apj, 350, 334

\bibitem[Henry and McCarthy(1993)]{HM93} Henry, T. J. and
McCarthy, D. W. Jr.  1993, \aj, 106, 773

\bibitem[Henry et al.(1997)]{hen97} Henry, T. J., Ianna, P. A., 
Kirkpatrick, J. D. and Jahreiss, H. 1997, \aj, 114, 388

\bibitem[Koerner et al.(1999)]{koern99} Koerner, D. W., Kirkpatrick, J. D., McElwain, M. W. and Bonaventure, N. R. 1999, \apjl, 526, L25

\bibitem[Luyten(1977)]{luyt} Luyten, W. 1977, Proper Motion Survey with the 48-Inch Schmidt Telescope (Minneapolis: University of Minnesota Press)

\bibitem[Lyot(1939)]{lyot39}Lyot, M. B. 1939, \mnras, 99, 578

\bibitem[Mart\'{\i}n et al.(1998)]{mart98} Mart\'{\i}n, E. L., 
Zapatero-Osorio, M. R. and Rebolo, R. 1998, ASP Conf.\ Ser., 134, 507

\bibitem[Matthews et al.(1996)]{M96} Matthews, K., Nakajima, T.,
Kulkarni, S. R. and Oppenheimer, B. R. 1996, \aj, 112, 1678

\bibitem[McCaughrean and Stauffer(1994)]{mcst94} McCaughrean,
M. J. and Stauffer, J. R. 1994, \aj, 108, 1383

\bibitem[Nakajima et al.(1994)]{N94} Nakajima, T., Durrance,
S. T., Golimowski, D. A. and Kulkarni, S. R. 1994, \apj, 428, 797

\bibitem[Nakajima et al.(1995)]{N95} Nakajima, T., Oppenheimer,
B. R., Kulkarni, S. R., Golimowski, D. A., Matthews, K. and Durrance,
S. T.  1995, \nat, 378, 463

\bibitem[Oppenheimer et al.(1995)]{O95} Oppenheimer, B. R.,
Kulkarni, S. R., Matthews, K. and Nakajima, T. 1995, Science, 270, 1478

\bibitem[Oppenheimer et al.(1998)]{O98} Oppenheimer, B. R.,
Kulkarni, S. R., Matthews, K. and van Kerkwijk, M. H.
1998, \apj, 502, 932

\bibitem[Oppenheimer et al.(2000)]{O00} Oppenheimer, B. R.,
Kulkarni, S. R. and Stauffer, J. R. 2000, in {\it Protostars and
Planets IV} (Tucson: University of Arizona Press, V. Mannings, A. Boss
and S. Russell, eds.)

\bibitem[Perryman et al.(1997)]{esa97} Perryman, M. A. C. et al. 1997,
\aap, 323, L49

\bibitem[Reid et al.(1995)]{JGthesis} Reid, I. N., Hawley, S. L. and Gizis, J. E. 1995, \aj, 110, 1838

\bibitem[Reid and Gizis(1997)]{reid97} Reid, I. N. and Gizis,
J. E. 1997, \aj, 113, 2246

\bibitem[Reid et al.(1999)]{reid99} Reid, I. N. et al. 1999, \apj, 521, 613

\bibitem[Schroeder et al.(2000)]{schr00} Schroeder, D. J. et al. 2000, \aj, 119, 906

\bibitem[Skrutskie et al.(1989)]{skrut89} Skrutskie, M. F, Forrest, W. J. and Shure, M. 1989, \aj, 98, 1409

\bibitem[Simons et al.(1996)]{sim96} Simons, D. A., Henry,
T. J. and Kirkpatrick, J. D. 1996, \aj, 112, 2238

\bibitem[van Altena et al.(1995)]{yale95}van Altena, W. F.,
Lee, J. T. and Hoffleit, D. 1995,{\it The General Catalog of Trigonometric 
Parallaxes, Fourth Edition} (New Haven, CT: Yale University Observatory)

\bibitem[van Biesbroeck(1961)]{vb61} van Biesbroeck, G. 1961, \aj, 66, 528

\bibitem[Weis(1984)]{weis84} Weis, E. W. 1984, \apjs, 55, 289

\end{thebibliography}
\end{document}